\documentclass{article}         
\usepackage[pdftex,dvipsnames]{xcolor}  
\usepackage{mathtools,bm,comment,amssymb,enumitem,hyperref,cite,array,float,subcaption,float,pgfplots,cancel}
\usepackage[capitalise,nameinlink]{cleveref}
\usepackage[left=.75in,right=.75in,top=.75in,bottom=.75in]{geometry}
\usepackage{makeidx,color,boxedminipage}
\usepackage{graphicx,float}
\usepackage{amsmath,amsthm,amsfonts,amscd,amssymb} 

\usepackage{xargs}                      

\usepackage{authblk} 

\usepackage{soul} 

\theoremstyle{definition}

\theoremstyle{remark}

\crefname{rem}{Remark}{Remarks}

\newcommand{\eqn}[1]{(\ref{#1})}

\newcommand{\bolds}[1]{\boldsymbol{#1}}
\newcommand{\bx}{\bolds{x}}
\newcommand{\bn}{\bolds{n}}

\newcommand{\bbR}{\mathbb{R}}
\newcommand{\kk}{\mathcal{K}}
\newcommand{\p}{\mathcal{P}}

\newcommand{\calG}{\mathcal{G}}
\newcommand{\calN}{\mathcal{N}}

\newcommand{\bphi}{\bolds{\phi}}
\newcommand{\phib}{\bar{\phi}}
\newcommand{\bVIF}{\bolds{VIF}}

\newcommand{\pyr}{\phib_{P}}
\newcommand{\lac}{\phib_{L}}
\newcommand{\pyrlac}{\bar{\bphi}}

\newcommand{\pyrh}{\phi_{P}}
\newcommand{\lach}{\phi_{L}}

\newcommand{\pyrvh}{\phi_{PV}}
\newcommand{\lacvh}{\phi_{LV}}

\newcommand{\sigp}{s_{P}}
\newcommand{\sigl}{s_{L}}
\newcommand{\sigpl}{\bolds{s}}

\newcommand{\testfn}{\tilde{\phi}}

\newcommand{\bA}{\bolds{A}}
\newcommand{\bC}{\bolds{C}}

\newcommand{\kth}[1]{${#1}^{\text{th}}$}

\newcommand{\bmatx}[1]{\begin{bmatrix}
    #1
\end{bmatrix}}
\newcommand{\weakDot}[2]{\left({#1}, {#2}\right)}

\DeclareMathOperator*{\argmax}{\mathrm{arg max}}

\newcommand{\SNR}{\mathrm{SNR}}

\title{Mutual-Information Based Optimal Experimental Design for Hyperpolarized  $^{13}$C-Pyruvate MRI\thanks{Corresponding authors: Prashant K. Jha (prashant.jha@austin.utexas.edu) and David T. Fuentes(dtfuentes@mdanderson.org)}}

\author[1]{Prashant K. Jha\thanks{prashant.jha@austin.utexas.edu}}
\author[2]{Christopher Walker\thanks{cmwalker@mdanderson.org}}
\author[2]{Drew Mitchell\thanks{DMitchell2@mdanderson.org}}
\author[1]{J. Tinsley Oden\thanks{oden@oden.utexas.edu}}
\author[2]{Dawid Schellingerhout\thanks{dawid.schellingerhout@mdanderson.org}}
\author[2]{James A. Bankson\thanks{jbankson@mdanderson.org}}
\author[2]{David T. Fuentes\thanks{dtfuentes@mdanderson.org}}

\affil[1]{Oden Institute for Computational Engineering and Sciences, The University of Texas at Austin, Austin, TX 78712, USA}
\affil[2]{Department of Imaging Physics, MD Anderson Cancer Center, Houston, TX 77320, USA}

\begin{document}                

\maketitle

\begin{abstract}
A key parameter of interest recovered from hyperpolarized (HP) MRI measurements
is the apparent pyruvate-to-lactate exchange rate, $k_{PL}$, for measuring tumor
metabolism. This manuscript presents an information-theory-based optimal
experimental design (OED) approach that minimizes the uncertainty in the rate
parameter, $k_{PL}$, recovered from HP-MRI measurements.
Mutual information (MI) is employed to measure the information content of the HP
measurements with respect to the first-order exchange kinetics of the pyruvate
conversion to lactate. Flip angles of the pulse sequence acquisition are
optimized with respect to the mutual information. Further, a spatially varying
model (high-fidelity) based on the Block  Torrey equations is proposed and
utilized as a control. 
A time-varying flip angle scheme leads to a higher parameter optimization that
can further improve the quantitative value of mutual information over a constant
flip angle scheme. However, the constant flip angle scheme leads to the best
accuracy and precision when considering inference from noise-corrupted data.
For the particular MRI data examined here, pyruvate and lactate flip angles of
35 and 28 degrees, respectively, were the best choice in terms of accuracy and
precision of the parameter recovery. Moreover, the recovery of rate parameter
$k_{PL}$ from the data generated from the high-fidelity model highlights the
influence of diffusion and strength of vascular source on the recovered rate
parameter. Since the existing pharmacokinetic models for HP-MRI do not account
for spatial variation, the optimized design parameters may not be fully optimal
in a more general 3D setting.
\end{abstract}

\section{Introduction}
The potential of hyperpolarized $^{13}$C-Pyruvate magnetic resonance imaging (HP-MRI) to 
characterize cancer biology, predict progression,
and monitor early responses to treatment is well known (e.g., ~\cite{nelson2013metabolic,
Bankson2015, kurhanewicz2019hyperpolarized, granlund2020hyperpolarized, miloushev2016hyperpolarization, miloushev2018metabolic, aggarwal2017hyperpolarized, gallagher2020imaging, woitek2020hyperpolarized}). Ongoing studies in
prostate, brain, breast, liver, cervical, and ovarian
cancer~\cite{kurhanewicz2019hyperpolarized, granlund2020hyperpolarized, nelson2013metabolic} have
shown that HP $^{13}$C-Pyruvate MRI is safe and effective. 
One of the central aspects of HP-MRI that make it appealing 
is the elevated chemical conversion of pyruvate to lactate, even in the presence of oxygen, via 
lactate dehydrogenase (LDH), also referred to as the Warburg effect \cite{warburg1956origin,
vander2009understanding}. 
The higher production of lactate has been shown to correlate with disease presence, 
the aggressiveness of the disease (e.g., cancer and inflation), and response to therapy. 
The rate of
pyruvate-to-lactate exchange ($k_{PL}$) is a crucial parameter of 
interest in locating aggressive disease and assessing the biological state of the
tissue.  HP-MRI presents a unique opportunity to observe tumor metabolism
\textit{in vivo}~\cite{Bankson2015, kurhanewicz2019hyperpolarized,
miloushev2016hyperpolarization, miloushev2018metabolic}
and use this information to make inferences about 
tumor aggressiveness and response to therapy.
However, a recent white
paper~\cite{kurhanewicz2019hyperpolarized} highlights the need for further
development of spatial, temporal, and spectral encoding strategies that minimize
uncertainty while maximizing the efficiency of HP-MRI. An example of 
uncertainty is the variability of the reported HP measurements in the
literature~\cite{grist2019quantifying,lee2020lactate,autry2019comparison}. In
the present work, we develop an information-theory-based approach 
to determine the optimal MRI design parameters, such as flip angles,
with a goal of reducing the uncertainty in the recovered rate parameter, $k_{PL}$. 

HP $^{13}$C-pyruvate MRI scans consist of several steps  \cite{kurhanewicz2019hyperpolarized,
miloushev2016hyperpolarization};  first, a mixture is prepared consisting of $^{13}$C-enriched metabolites ($[1-^{13}$C$]$-pyruvate) and an electron paramagnetic agent (EPA) that mediates polarization buildup and transfer. 
Using the Dynamic Nuclear Polarization (DNP) method, one of the atoms of the $^{13}$C-pyruvate molecule is polarized heavily; $[1-^{13}$C$]$-pyruvate means that the first carbon is labeled with the $^{13}$C isotope, which is visible by magnetic resonance (MR). 
On the other hand, 98.9 percent of the carbon are $^{12}$C which has a spin quantum number of zero and therefore can not be polarized or detected by MR. 
Once the mixture is processed to obtain the hyperpolarized pyruvate, the temperature of the solution is brought to the body temperature and quickly injected into the patient. 
Following the injection, the patient is scanned using a gradient echo pulse sequence to measure the temporal evolution of the $^{13}$C-pyruvate and $^{13}$C-lactate (lactate produced from hyperpolarized pyruvate too are visible in the scan) magnetizations. 
The artificial polarization of $^{13}$C-pyruvate (and $^{13}$C-lactate) decays rapidly, and therefore the injection and subsequent MR scans must be performed quickly, within $60 - 120$ seconds. 
One of the critical reasons for using $^{13}$C-pyruvate is its compatibility; it is an endogenous substance that plays a central role in catabolic and anabolic metabolism. 
The repetition time of the MR pulse sequence is limited by the number of slices and relaxation (decay) of polarized molecules, typically above $0.5$ and below $5 - 10$ seconds. 
The key scan parameter at each scan is the flip angle (or excitation angle); a larger flip angle provides a higher signal (the magnetization of pyruvate and lactate) intensity; however, the signal that remains for the following scan is reduced. 
Therefore, the choice of flip angles can impact the quality of data and, thus, the quality of the recovered parameter. 
There are other design parameters that can be considered, for example, spectral encoding scheme, gradient trajectory, etc. 
However, in this work, the flip angles for pyruvate and lactate are chosen as design parameters. 

The time history of pyruvate and lactate magnetization within the imaging voxels constitutes the data of interest. 
Together with a pharmacokinetic HP-MRI model, pyruvate and lactate magnetization data are employed to recover the pyruvate-to-lactate apparent exchange rate, $k_{PL}$; e.g., \cite{Bankson2015, walker2019effects, maidens2015optimal}. 
Accuracy and uncertainty in the recovered rate parameter depend on the data's information content and the pharmacokinetic model's fidelity. 
This work aims to determine the MRI design parameters that increase the information content in the data and reduce the uncertainty in the rate parameter. 
The mutual information (MI) between the data and critical model parameters is utilized toward this goal. 
Information theory and, specifically, mutual information provides a rigorous mathematical framework to identify optimal MR imaging data and acquisition parameters that maximize image reconstruction quality. 
It offers a quantitative methodology for optimizing acquisition protocols to improve reproducibility. 

To verify the reduction in uncertainty of recovered rate parameter $k_{PL}$ using optimal design parameters, 
we generate synthetic data from low-fidelity pharmacokinetic and high-fidelity models. 
The high-fidelity model is obtained using continuum mixture theory and includes two linear coupled partial differential equations for spatially varying HP pyruvate and lactate in the tissue. 
It is shown that for data from both models, optimal design parameters reduce uncertainty in the pyruvate-to-lactate apparent exchange rate, $k_{PL}$.

The article is structured as follows: In \cref{s:relwork} related work is reviewed. 
\cref{s:model} presents the  pharmacokinetic and high-fidelity
models used to produce data for verification of optimal scan parameters. 
The optimization of scan parameters based on 
mutual information is discussed in \cref{s:mi}. The results obtained from the proposed model 
are presented in \cref{s:results}. 
In \cref{s:disc} some conclusions are drawn, and 
future directions are proposed. 
The codes and relevant files to reproduce the results will be publicly available in the following GitHub repository:
\texttt{https://github.com/prashjha/HyperpolarizedMRI}.

\section{Related works}\label{s:relwork}
Formulating optimal HP acquisition strategies is challenging and tightly coupled with the analysis of appropriate models of MRI physics. 
Semi-quantitative metrics such as the ratio of the integrals of the lactate-to-pyruvate signals (area under the curve, AUC)  are often preferred due to their simplicity. However, the  lactate-to-pyruvate ratio is biased by HP pyruvate in blood vessels that are inaccessible to enzymes that mediate conversion from HP pyruvate to lactate.  For example, if the vascular blood volume decreases by 10 percent in a region of a tumor, the lactate-to-pyruvate ratio could increase even if the true metabolic state of cells does not change. The lactate-to-pyruvate ratio is also affected by the excitation scheme. Small flip angles consume less pyruvate magnetization and permit the signal pool to remain longer for the conversion to lactate.  The potential for variability is realized in the  literature. In applications to brain cancer (glioma) Grist~\cite{grist2019quantifying} reports a lactate to pyruvate ratio of .25 $\pm$ .08 and .22 $\pm$ .06 in white matter and gray matter, respectively. Lactate-to-pyruvate ratios greater than 1.0 are reported in \cite{lee2020lactate} in gray matter.
Lactate-to-pyruvate ratios in white matter of 0.43 $\pm$ 0.14 were reported in  \cite{autry2019comparison}.

The information theoretic approach developed in this work is an extension of
optimal experimental design approaches that use the Fisher information matrix and the
Cramer -- Rao bound as a lower bound on the variance of unbiased
estimators~\cite{maidens2015optimal,maidens2016optimizing,marseille1996bayesian,brihuega2003optimization,poot2010optimal,cercignani2006optimal,reeves1999sequential,ji2008bayesian,seeger2010optimization}. Indeed, the de Bruijn identity~\cite{Cover2012} provides a direct connection between derivatives of our entropy calculations and the Fisher information matrix.
However, optimizing the Fisher information matrix requires estimates of the unknown
tissue parameters, such as $T_1$ relaxation losses and pyruvate-to-lactate
conversion rate (that we are trying to recover), to calculate the Fisher
information.  The Fisher information must be iteratively re-optimized as more
accurate estimates of the tissue properties are obtained.  In contrast, the present
approach provides a mathematical framework to directly include the tissue
parameter uncertainty in the mutual information and considers a range of tissue
parameters (determined by the input probability distributions) to calculate and
optimize the mutual information.

\section{Hyperpolarized (HP) MRI model}\label{s:model}
Consider a tissue domain within which different constituents evolve depending on the 
local environment which includes, for example, interstitial, vascular, and cellular spaces. 
In this work, following \cite{walker2019effects, Bankson2015}, two spatial compartments, namely, interstitial and vascular, each containing HP pyruvate and lactate and complement of these two constituents, are considered. 
First, the spatially invariant pharmacokinetic model developed in \cite{walker2019effects} is presented. 
The model accounts for $T_1$ relaxation loss, signal loss due to excitation at each scan, 
and pyruvate-to-lactate and lactate-to-pyruvate exchanges. 
This model is referred to as the low-fidelity (LF) model for convenience. 
To generate synthetic data as a control to test optimal design parameters, 
the low-fidelity model is extended using principles of continuum mixture theory to 
obtain a PDE-based model, 
referred to as the high-fidelity (HF) model.  
The HF model includes the vascular-tissue exchange in an attempt to produce a more accurate 
perfusion of HP pyruvate. 
The two models are presented next. 

\subsection{Spatially invariant  HP-MRI low-fidelity (LF) model}\label{ss:model}
A two compartment HP-MRI model (interstitial and vascular) consisting of two constituents, HP pyruvate and lactate, is considered. 

Let $\pyr = \pyr(t)$ and $\lac =\lac(t)$ denote the average volume fraction of HP
pyruvate and lactate, respectively, over the tissue domain $\Omega \subset \bbR^3$ at time $t \in [0,T]$. 
Discrete times at $N$ scans are denoted by $t_k$, $1\leq k \leq N$; $\theta^k_P$
and $\theta^k_L$ are flip angles in \kth{k} scan whereas $TR_k = t_k - t_{k-1}$,
$k > 1$, are repetition times and $TR_1 = 0$. 
With $\pyrlac = (\pyr, \lac)^T$, the HP pyruvate and lactate available for measurement at 
the \kth{(k+1)} scan, $k \geq 1$, are given by \cite{walker2019effects}
\begin{equation}\label{eq:model}
\pyrlac(t_{k+1}) = \exp\left[TR_{k+1} \bA \right] \bC^k \pyrlac(t_k) + \frac{ k_{ve}}{\nu_e} \int_{t_k}^{t_{k+1}} \exp\left[(t_{k+1} - \tau) \bA \right] \bVIF(\tau) d\tau,
\end{equation}
where the matrix $\bA$ accounts for $T_1$ relaxation losses and pyruvate-lactate exchanges
\begin{align}\label{eq:A}
\bA = \bmatx{-\frac{1}{T_{1,P}} - k_{PL} -\frac{ k_{ve}}{\nu_e}  & k_{LP} \\
k_{PL} & -\frac{1}{T_{1,L}} - k_{LP}} .
\end{align}
Here $T_{1,P}, T_{1,L}$ denote the $T_1$ relaxation times (s), $k_{PL}, k_{LP}$ pyruvate-to-lactate and lactate-to-pyruvate exchange rates (s$^{-1}$), $k_{ve}$ vascular-tissue exchange rate (s$^{-1}$), and $\nu_e$ the extravascular volume fraction.  In \eqref{eq:model}, $\bC^k$ denotes the matrix that takes into account the loss of signal due to excitation at \kth{k} scan:
\begin{align}\label{eq:C}
\bC^k = \bmatx{\cos(\theta_P^k) & 0 \\
0 & \cos(\theta_L^k)} .
\end{align}
Lastly, $\bVIF = \bVIF(t)$ is the vascular input function assumed to take the form:
\begin{align}
\bVIF(t) = \bmatx{\sigma_P \gamma(t - \bar{t}_0, \alpha_P, \beta_P) \\ 0},
\end{align}
where $\sigma_P, \alpha_P, \beta_P$ are constants, and $\gamma$ denotes a gamma probability density function given by
\begin{equation}\label{eq:gamma}
\gamma(t, a, b) = \frac{1}{b^a \Gamma(a)} t^{a-1} \exp\left[-\frac{t}{b}\right] .
\end{equation}
The constant $\bar{t}_0$ is associated with bolus arrival time 
and is treated as one of the uncertain model parameters. 

In \eqref{eq:model}, the parameters can be gathered in two different classes: 1) model parameters that depend on the tissue domain 
and includes $\p = (T_{1,P}, T_{1,L}, k_{PL}, k_{LP}, k_{ve}, \nu_e, \bar{t}_0)$, 
and 2) HP MRI design parameters such as repetition times and flip angles 
$\kk = (\{TR_i\}_{i=2}^N, \{\theta_P^i\}_{i=1}^N, \{\theta_L^i\}_{i=1}^N)$. 
For simplicity, the parameters $T_{1,P}, T_{1,L}, \nu_e, k_{LP}$ are assumed to be known and fixed so that $\p = (k_{PL}, k_{ve}, \bar{t}_0)$. 
Repetition times $TR_k$ are also considered fixed so that $\kk = (\{\theta_P^i\}_{i=1}^N, \{\theta_L^i\}_{i=1}^N)$.

\subsubsection{The total signal}
For simplicity, let $\pyrlac^k = \pyrlac(t_k)$, $\pyr^k = \pyr(t_k), \lac^k = \lac(t_k)$, where
$k = 1, 2, ..., N$. 
Suppose prior to the \kth{k} scan, the available HP pyruvate
signal is $\pyr^k$. After taking the \kth{k} scan,
due to signal excitation loss, $\cos(\theta_P^k) \pyr^{k}$ will be available for the next measurement.
A similar argument can be made for $\lac^k$. 
Further, the signals measured at the \kth{k} scan are, see \cite{walker2019effects},
\begin{align}\label{eq:signal}
	\sigpl^k =  \bmatx{\sigp^k \\ \sigl^k} &= \bmatx{\sin(\theta_P^k) & 0 \\ 0 & \sin(\theta_L^k)} \left(\nu_e \pyrlac^{k} + (1-\nu_e) \bVIF(t_{k}) \right) = \bmatx{\sin(\theta_P^k)\left(\nu_e \pyr^k + (1-\nu_e) \bVIF_1(t_k) \right) \\ \sin(\theta_L^k)\left(\nu_e \lac^k + (1-\nu_e) \bVIF_2(t_k) \right)}
\end{align}
i.e., one only measures the $\sin(\theta_P^k)$ and $\sin(\theta_L^k)$ fractions of the magnetization leaving the $\cos(\theta_P^k)$ and $\cos(\theta_L^k)$ fractions for the next measurement. The total signal is the sum of the individual signals at different scans and is given by
\begin{equation}\label{eq:G}
	\calG = \calG(\kk, \p) = \sum_{k=1}^N \left( \sigp^k + \sigl^k \right),
\end{equation}
where the dependence of $\calG$ on the design parameters $\kk$ and model parameters $\p$ 
is clear from \eqref{eq:model} and \eqref{eq:signal}.

\subsection{HP-MRI high-fidelity (HF) model for data generation}\label{ss:modelhf}
The verification of uncertainty reduction in recovered $k_{PL}$ on living subjects
using the optimal design parameters 
is a challenging undertaking. 
One approach for verification is to use the synthetic data generated from the computational models. 
Here, the two models are employed to generate synthetic data; 
the first model is the same ODE model described in previous section. 
As a second model, somewhat more realistic model that includes spatial variation of 
the HP pyruvate and lactate is developed using the principles of continuum mixture theory 
and invoking the principle of conservation of mass; see \cite{oden2010general,riviere2008discontinuous,peacemanfundamentals,faust1979geothermal}.

Let $\Omega \subset \bbR^3$ be an open bounded domain representing a tissue. At any $\bx \in \Omega$, 
consider an RVE (Representative Volume Element at sub-tissue-scale ($\approx$ 1 mm). 
At this scale, a typical RVE may include three compartments: \textit{interstitial}, \textit{vascular}, and \textit{cellular}. 
These compartments may host various constituents, e.g., proteins, plasma, etc., 
and constituents can interact with other and exchange  mass at the interfaces between compartments. 
Of particular interest in this work are the HP pyruvate and lactate in interstitial and vascular compartments. 
Following some simplifying assumptions outlined in \cref{s:derivationPDEModel}, the evolution of pyruvate and lactate volume fractions, $\phi_P = \phi_P(t, \bx)$ and $\phi_L = \phi_L(t, \bx)$, $t\in (0, t_F)$, $\bx \in \Omega$, in interstitial compartment can be described by a coupled system of reaction-diffusion partial differential equations. HP agents in vascular compartments, denoted by $\phi_{PV} = \phi_{PV}(t, \bx)$ for HP pyruvate and $\phi_{LV} = \phi_{LV}(t, \bx)$ for lactate, $t \in (0, t_F), \bx \in \Omega$, are assumed to be fixed and given as a function of time and space. 

To account for the instantaneous loss of signal (or volume fractions) of HP agents at discrete scan points $t_k$, $1\leq k\leq N$, consider $\phi^k_A = \phi^k_A(t, \bx)$, where $t\in [t_{k-1}, t_k)$, $\bx \in \Omega$, $A\in \{P, L, PV, LV\}$. 
The balance of mass for constituents in interstitium based on simplifying assumptions can be expressed in differential form, for $t\in (t_{k-1}, t_k), \bx \in \Omega$, as follows
\begin{equation}\label{eq:modelHFRep}
	\begin{split}
		\frac{\partial }{\partial t} \pyrh^k &= \nabla \cdot \left( D_P \nabla \pyrh^k \right)  + S_P(\pyrh^k, \lach^k) + J_P(\pyrh^k, \pyrvh^k),  \\
	\frac{\partial }{\partial t} \lach^k &= \nabla \cdot \left( D_L \nabla \lach^k \right)  + S_L(\pyrh^k, \lach^k) + J_L(\lach^k, \lacvh^k),
	\end{split}
\end{equation}
where $D_P, D_L$ are diffusivities of pyruvate and lactate, $S_P = S_P(\pyrh^k,
\lach^k), S_L = S_L(\pyrh^k, \lach^k)$ volume source terms, and $J_P =
J_P(\pyrh^k, \pyrvh^k), J_L = J_L(\lach^k, \lacvh^k)$  vascular-interstitial volume
exchange terms. 
The initial condition for the HP pyruvate and lactate, $\phi_P^k, \phi^k_L$, should take into account the loss of signal due to excitation in the \kth{(k-1)} scan. 
For example, HP pyruvate just before the \kth{(k-1)} scan is $\phi^{k-1}(t_{k-1})$, and the available signal due to instant loss of signal after \kth{(k-1)} scan is $\cos(\theta^{k-1}_P) \phi^{k-1}_P(t_{k-1}) $. 
Based on the above argument, initial conditions are taken as
\begin{equation}\label{eq:icHF}
	\pyrh^k(t_{k-1}, \bx) = \cos(\theta_P^{k-1}) \pyrh^{k-1}(t_{k-1}, \bx), \quad \lach^k(t_{k-1}, \bx) = \cos(\theta_L^{k-1})\lach^{k-1}(t_{k-1}, \bx), \qquad \forall \bx \in \Omega, 1\leq k \leq N .
\end{equation}
Homogeneous Neumann boundary conditions are assumed to hold, i.e.,
\begin{equation}
	\nabla \pyrh^k(t, \bx) \cdot \bn = 0, \quad \nabla \lach^k(t, \bx) \cdot \bn = 0, \qquad \forall t\in [t_{k-1}, t_k], \bx \in \partial \Omega,
\end{equation}
where $\bn$ is the unit outward normal of the boundary $\partial \Omega$. For the well-posedness of \eqref{eq:icHF} for $k=1$, it is assumed that $\phi^0_P(t_0, \bx) = \phi^0(t_0, \bx) = 0$, for all $\bx \in \Omega$, and $\theta^0_P = \theta^0_L = 0$. 

The volume source terms $S_P, S_L$ are defined as follows:
\begin{align}\label{eq:Sterms}
S_P(\pyrh^k, \lach^k) &= \left(-\frac{1}{T_{1,P}} - k_{PL}\right) \pyrh^k + k_{LP} \lach^k, & S_L(\pyrh^k, \lach^k) &= \left(-\frac{1}{T_{1,L}} - k_{LP}\right) \lach^k + k_{PL} \pyrh^k,
\end{align}
and the tissue-vascular volume exchange terms are simply taken as the volume sources restricted on vascular subdomain, $\Omega_V \subset \Omega$, i.e.,
\begin{align}\label{eq:Jterms}
	J_P(\pyrh^k, \pyrvh^k) &= L_P \pyrvh^k \chi_{\Omega_V}, &J_L(\lach^k, \lacvh^k) &= L_L \lacvh^k \chi_{\Omega_V},
\end{align}
where $\chi_A$ is the characteristics function of set $A$, i.e., $\chi_A(\bx) = 1$ if $\bx \in A$ and $\chi_A(\bx) = 0$ if $\bx \notin A$, and $L_P, L_L$ permeabilities. 

HP agents in the vascular compartment are fixed as, for all $t\in [t_k, t_{k+1})$,
\begin{align}\label{eq:vifvasc}
	\pyrvh^k(t, \bx) = 
		\begin{cases}
			\sigma_P \gamma(t - \bar{t}_0, \alpha_P, \beta_P), &\qquad \text{if } \bx \in \Omega_V, \\
			0, &\qquad \text{otherwise}
		\end{cases} 
\end{align}
and $\lacvh^k(t, \bx) = 0$, $\forall \bx \in \Omega$. Here $\bar{t}_0, \sigma_P, \alpha_P, \beta_P$ are constants that control the magnitude and shape of the gamma function $\gamma$ defined in \eqref{eq:gamma}. 

Next, the model is presented. Further details about the setup of this model and simplifying assumptions are provided in \cref{s:derivationPDEModel}. 

\subsubsection{Signal from the high-fidelity model}
The magnetization intensity of constituent $A$, $A\in \{P, L, PV, LV\}$, at $\bx \in \Omega$ at \kth{k} scan is assumed to be proportional to the volume fraction of the constituent, $\phi^k_A$, 
i.e., there are constants $C_A$, $A\in \{P, L, PV, LV\}$, such that $C_A \sin(\theta^k_A) \phi^k_A(t_k, \bx)$ is the measured signal of constituent $A$ at the \kth{k} scan. 
Here, $\theta^k_{PV} = \theta^k_{P}$ and $\theta^k_{LV} = \theta^k_{L}$, where $\theta^k_P, \theta^k_L$ are the flip angles. Without loss of generality, it is assumed that $C_A = 1$, for $A\in \{P, L, PV, LV\}$. 

\section{Mutual information based optimization of MR scan parameters}\label{s:mi}
A major goal of this study is to formulate a optimization problem to determine the design 
parameters $\color{black}\kk = ( \{\theta_P^i\}_{i=1}^N, \{\theta_L^i\}_{i=1}^N)$  such that the MRI measurements reduce uncertainty in the rate parameter, $k_{PL}$. 
Treating total signal, $\calG$ defined in \eqref{eq:G}, as the data, and model
parameters, $\color{black}\p = (k_{PL}, k_{ve}, \bar{t}_0)$, together with data as random variables, an optimization problem of maximizing the mutual information (MI) between the data and model parameters is proposed. 
It is shown that uncertainty in recovered $k_{PL}$ from synthetic noisy data is reduced when optimal design parameters are considered; see \cref{s:results}. 

In what follows, after defining some preliminary quantities, 
the mutual information between data and the model parameters is defined.
Let $z\in Z = \bbR$, $\p \in \Theta \subset \bbR^3$, and $\kk \in D
\subset \bbR^{2N}$, where $Z, \Theta, D$ are data, model parameter, and
design parameter spaces, respectively. 

\subsection{Prior, likelihood, and evidence}
Suppose $p_0(\p)$ is the prior probability distribution function (PDF)  of model parameters $\p$ and $p(z)$ is the PDF of the data $z$. Then the joint PDF $p(z, \p)$ must satisfy
\begin{equation}\label{eq:jointPDF}
p(z, \p) = p(z| \p) p_0(\p) = p(\p | z) p(z),
\end{equation}
where $p(z|\p)$ is the conditional PDF of data 
conditioned on model parameters $\p$ (also referred to as the {\it likelihood} function) and $p(\p| z)$ the conditional PDF of model parameters $\p$ conditioned on data $z$ ({\it posterior} of $\p$). 
The prior is taken as multi-variate Gaussian with mean $\hat{\p}$ and covariance matrix $\Sigma_\p$:
   \begin{equation}\label{eq:prior}
      \p \sim p_0(\p) = \calN(\hat{\p}, \Sigma_\p)
                      =  \frac{1}{2 \pi  \det{\Sigma_\p}} \exp\left( 
          - \frac{1}{2} \|\hat{\p}  - \p \|^2_{\Sigma^{-1}_\p}
                                                                  \right) .
   \end{equation}
Here $\|\hat{\p}  - \p \|^2_{\Sigma^{-1}_\p} = (\hat{\p}  - \p)\cdot \Sigma^{-1}_\p(\hat{\p}  - \p)$. Within this
Bayesian setting,~$\Sigma_p$, is representative of biological variability. 

To derive the expression for the likelihood function, first suppose that $\calG = \calG(\kk, \p)$ is the model prediction of data. 
Data and the model prediction are assumed to be related as follows:
\begin{equation}\label{eq:noise}
z = \calG(\kk, \p) + \varepsilon \qquad \Rightarrow \quad z - \calG(\kk, \p) \sim \calN(0, \sigma_z) ,
\end{equation}
where an additive model of noise is assumed and the noise, $\varepsilon$, is taken as Gaussian with a zero mean and standard deviation $\sigma_z$, i.e., $\varepsilon \sim \calN(0, \sigma_z^2)$. 

Therefore, the likelihood function $p(z|\p)$ takes the Gaussian form: 
  \begin{equation} \label{eq:gaussian}
      p(z|\p)   =  \calN(\calG(\kk,\p),\sigma_z)  
                      =  \frac{1}{2 \pi \sigma_z} \exp\left( - \frac{1}{2\sigma_z^2} \|\calG(\kk,\p)  - z \|^2\right) .
  \end{equation} 
Here $|| \cdot ||$ denotes the Euclidean norm. Technically, $p(z| \p)$ is also conditioned on $\kk$, but, for simplicity in notation, the dependence on $\kk$ is suppressed. 

With $p_0(\p)$ and $p(z|\p)$ defined as above, the joint PDF $p(z, \p)$ can be computed using \eqref{eq:jointPDF}. Further, using \eqref{eq:jointPDF}, the PDF of data $z$ (evidence), $p(z)$, can be computed by marginalizing $p(z, \p)$ with respect to $\p$:
\begin{equation}\label{eq:evid}
p(z) = \int_{\Theta} p(z, \p) d\p = \int_{\Theta} p(z| \p) p_0(\p) d \p,
\end{equation}
where $\Theta$ is the parameter space. 

\subsection{Mutual information}
Next, the statistical mutual information is defined and the optimization problem for design parameters $\kk$ is posed. 
Given HP-MRI data, the accurate inference of pyruvate-to-lactate exchange rate
parameters (and other parameters in $\p$) depends on the specific choice of
control (design) parameters $\kk$ as selection of $\kk$ affects the information content in the measured data. 
The notion of mutual information~\cite{Cover2012} provides a way to quantify the 
information content about the model parameters $\p$ in the data $z$. 
The mutual information
between the two random variables $z$ and $\p$ with PDFs
$p(z)$ and $p_0(\p)$ and the joint PDF $p(z,\p)$ is defined as
\begin{equation} \label{mi}
I = I(\kk) = \int_Z \int_\Theta p(z, \p)\ln\left(\frac{p(z, \p)}{p_0(\p)p(z)}\right)d\p dz .
\end{equation}
Here, the mutual information $I$ depends only on design parameters $\kk$ and the forward model \eqref{eq:model}.

Using Bayes theorem, $p(z, \p) = p(z|\p)p_0(\p)$, it can easily be shown that
\begin{equation} \label{mi2}
I(\kk)=\int_Z\int_\Theta p(z|\p)p_0(\p)\ln\left(\frac{p(z|\p)p_0(\p)}{p_0(\p)p(z)}\right)d\p dz,
\end{equation}
or,
\begin{equation} \label{mi3}
I(\kk)=\int_Z\int_\Theta p(z|\p)p_0(\p)\ln\left[p(z|\p)\right]d\p dz - \int_Z p(z) \ln p(z)dz
       = H(z; \kk) - H(z|\p; \kk) ,
\end{equation}
where the second term in the above equation is identified as information entropy $H(z; \kk)$ and the first term as negative of the cross-information entropy, $H(z|\p; \kk)$. That is
\begin{equation}
H(z; \kk) = - \int_Z p(z) \ln p(z)dz, \qquad H(z | \p; \kk) = -\int_Z\int_\Theta p(z|\p)p_0(\p)\ln\left[p(z|\p)\right]d\p dz .
\end{equation}

\paragraph{Optimization problem}
In order to maximize the reduction in the uncertainty in the model parameters (i.e. to have the most confident estimates of the parameters $\p$), we propose to maximize the mutual information between the observation data and parameters of interest:
\begin{equation}\label{mimax}
\max_{\kk \in D} I(\kk) = I(\kk^\ast),
\end{equation}
where $\kk^\ast$ is the design parameter for which $I$ is maximum (assuming $\kk$ exist).

Given a Gaussian prior and conditional probability of the data with respect to the parameters, the entropy of the data conditioned on the model parameter, i.e., $H(z| \p)$,  is constant. Therefore, the optimization problem simplifies to
\begin{equation}\label{mimaxHz}
\kk^\ast = \argmax_{\kk \in D} I(\kk) = \argmax_{\kk \in D} H(z; \kk). 
\end{equation}

\subsection{Approximation of mutual information and information entropy}
Mutual information calculations are computationally demanding due to high-dimensional integration over the parameter and data spaces. Combinations of both quadrature and sampling methods have been employed for mutual information calculations, each of which is well-suited to certain function classes~\cite{Gerstner1998,Niederreiter1992,robert2013monte,gordon1996computer,Sloan1994,Genz1987,VanDooren1976,Barron1994,Cavagnaro2010,Drovandi2014,Ryan2014,Ryan2003,Ryan2016}. These methods include Monte Carlo and Quasi-Monte Carlo methods~\cite{Niederreiter1992,robert2013monte,gordon1996computer}, lattice rules~\cite{Sloan1994}, adaptive subdivision~\cite{Genz1987,VanDooren1976}, neural network approximations~\cite{Barron1994} and numerical quadrature~\cite{Cavagnaro2010}. 
Here the problem structure is used to accelerate computations and facilitate tractable numerical integration.
Following~\cite{mitchell2020information}, Gauss-Hermite quadrature is applied in each
dimension of mutual information integrals defined in \eqref{mi3} to numerically integrate multi-variate Gaussian random variables.
Quadrature order is increased until convergence is observed.

\section{Numerical results}\label{s:results}
In this section, the main results of our analysis are presented. First, 
the optimal design parameters for different signal-to-noise ratios (SNRs) are shown. 
{\color{black}Optimal design parameters for both temporally constant and varying
flip angles at each data acquisition are considered.}
Next, the reduction in uncertainty of $k_{PL}$ when using optimal design parameters 
generated synthetic data is shown; synthetic data are generated using the two models discussed in \cref{s:model}. 

For the optimization of mutual information and inverse problem to recover model
parameters from the noisy data, the automatic differentiation feature in MATLAB is
utilized.
Details on the numerical solution of 
the low and high-fidelity models are given in \cref{s:discretization}. 
The values of model parameters and MR scan (design) parameters are listed in Tables
\ref{tab:modelParam} and \ref{tab:designParam}, respectively. 

\begin{table}[h!]
\begin{center}
\begin{tabular}{|c|c|l|} 
\hline
\textbf{Parameter} & \textbf{Value} & \textbf{Description} \\
\hline
\hline
$L_P$, $L_L$ & $0.2$ s$^{-1}$ & Permeabilities for vessel-tissue mass exchange (only for HF model)\\
$D_P$, $D_L$ & $20$ cm$^2$/s$^{-1}$ & Diffusivities (only for HF model) \\
$T_{1,P}, T_{1,L}$ & \color{black} $30$ s, $25$ s & Relaxation times \\
$k_{PL}$ & $0.15$ s$^{-1}$ & Pyruvate-to-lactate apparent exchange rate\\
$k_{LP}$ & $0$ s$^{-1}$ & Lactate-to-pyruvate apparent exchange rate\\
$\bar{t}_0$ & \color{black}$4$ s & Bolus arrival time for HP pyruvate \\
$\sigma_P, \alpha_P, \beta_P$ & \color{black}$100$, $2.5$, $4.5$ s & Parameters in the gamma function \eqref{eq:vifvasc} \\
$k_{\nu_e}$ & 0.05 s$^{-1}$ & vascular-tissue exchange rate (only for LF model)\\
$\nu_e$ & 0.95 & extravascular volume fraction (only for LF model)\\
\hline
\end{tabular}
\end{center}
\caption{Default model parameters, $\mathcal{P}$, and remaining fixed model parameters used in initialization,
optimization, and verification steps. 
Some parameters are specific to either LF (low-fidelity) or HF (high-fidelity) model 
while the rest are the same in the two models. 
The parameters such as $D_P, D_L, L_P, L_L$ for HF model are assumed arbitrary values without loss of generality. 
}\label{tab:modelParam}
\end{table}

\begin{table}[h!]
\begin{center}
\begin{tabular}{|c|c|l|} 
\hline
\textbf{Parameter} & \textbf{Value} & \textbf{Description} \\
\hline
\hline
$N$          & \color{black}$30$ & Number of HP MRI scans\\
$TR_k$       & \color{black}$3$ s & Repetition times (for $k> 1$) \\
$\theta^k_P$ & \color{black}$20$ degrees & Flip angles for HP pyruvate (for all $k$) \\
$\theta^k_L$ & \color{black}$30$ degrees & Flip angles for HP lactate (for all $k$) \\
\hline
\end{tabular}
\end{center}
\caption{Default design parameters, $\mathcal{K}$, used in initialization,
optimization, and verification steps.}\label{tab:designParam}
\end{table}

\subsection{Optimized design parameters}\label{ss:oedRes}
\begin{figure}[h]
\centering
\begin{tabular}{ccc}
 \includegraphics[width=0.33\textwidth]{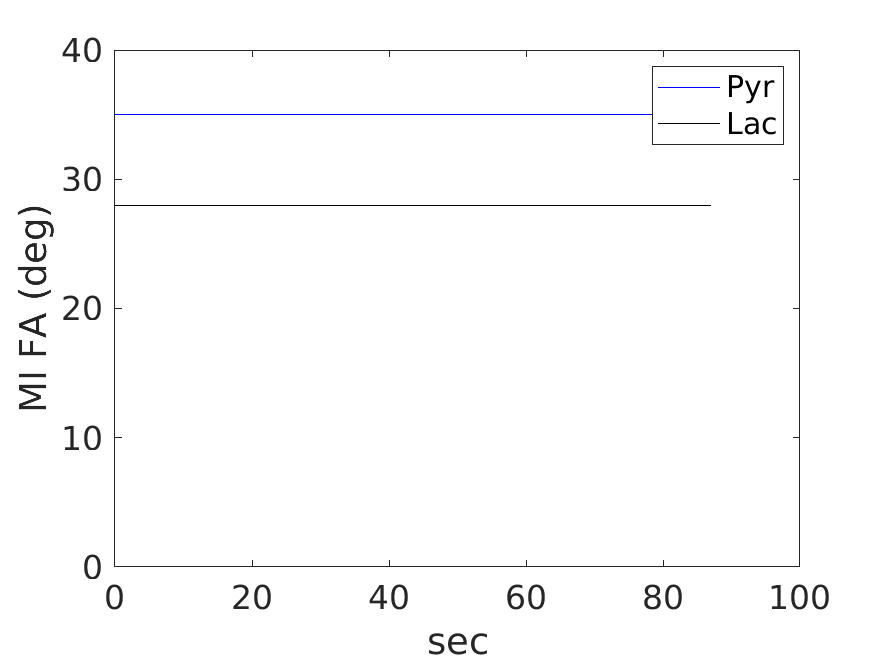}&
 \includegraphics[width=0.33\textwidth]{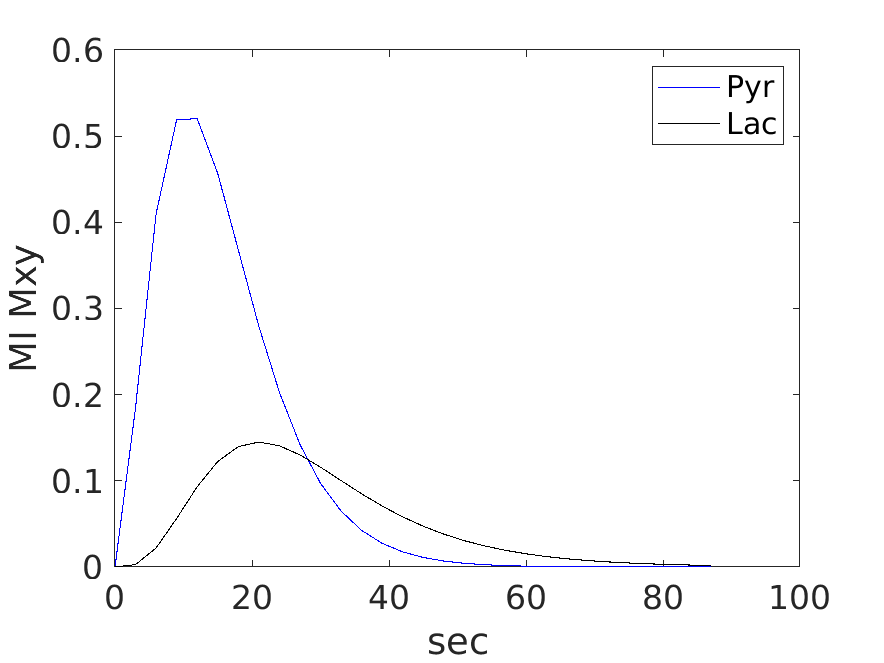}&
 \includegraphics[width=0.33\textwidth]{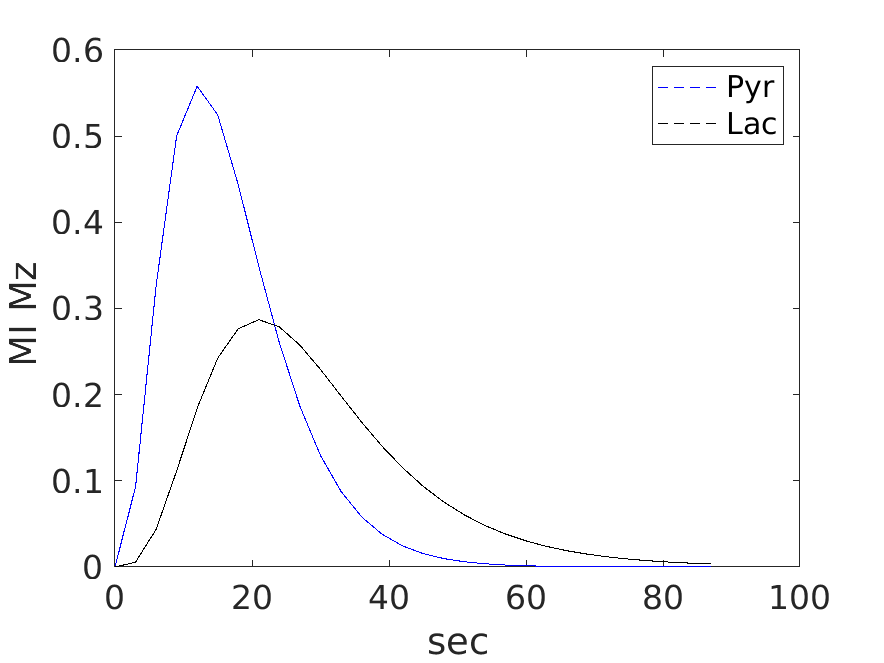}\\
(a) & (b) & (c) \\
 \includegraphics[width=0.33\textwidth]{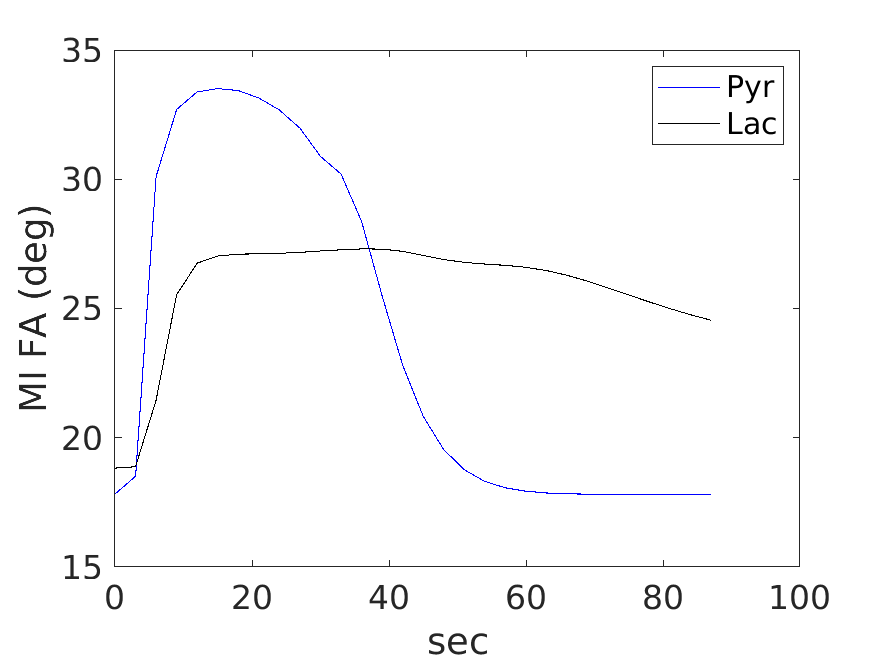}&
 \includegraphics[width=0.33\textwidth]{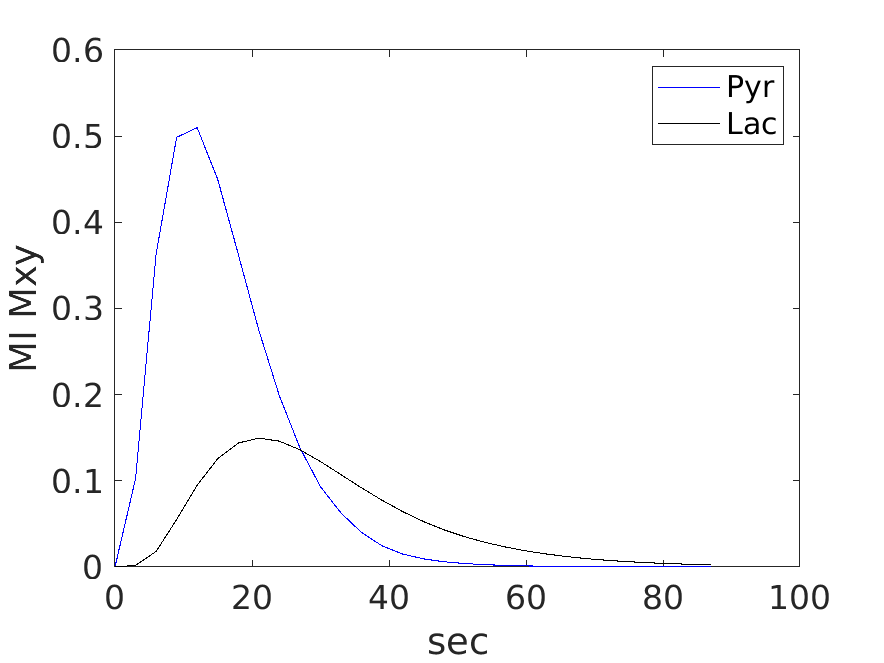}&
 \includegraphics[width=0.33\textwidth]{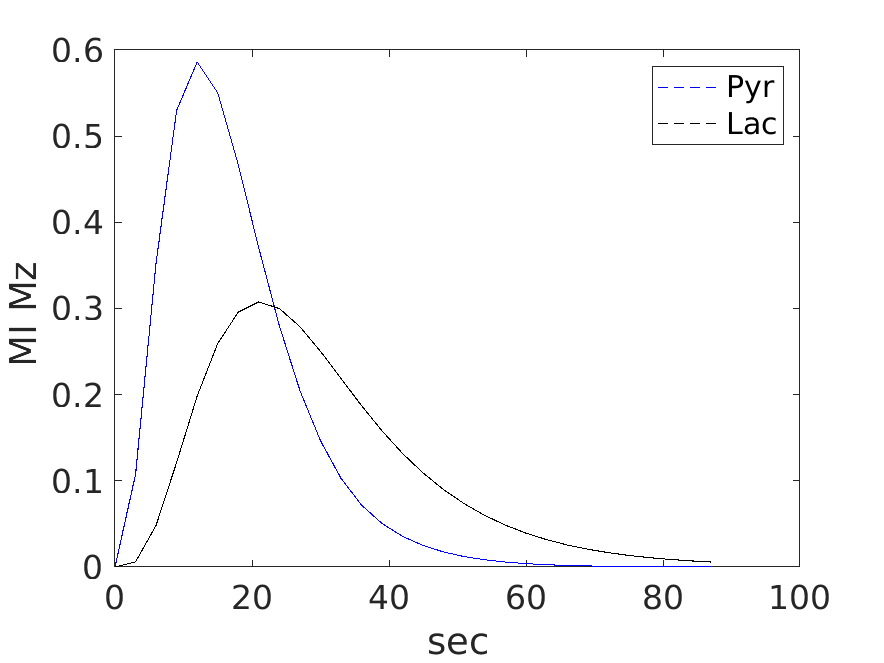}\\
(d) & (e) & (f) \\
\end{tabular}
\caption{Optimized design parameters along with the solution of the forward (LF) 
model for noise $\sigma_z(2)$, i.e., $SNR = 2$. The optimized flip angle scheme is
shown for  (a) constant flip angles throughout the acquisition (optimal angles are $\theta^k_P = 35$ degrees and $\theta^k_L = 28$ degrees, for all $1\leq k\leq N$) and (d)
allowing the flip angle to vary at each data acquisition.
The corresponding signal evolution of the transverse magnetization \eqn{eq:signal}
is shown in (b) for the constant flip angle scheme and (e) for
varying flip angles. Similarly, the longitudinal magnetization \eqn{eq:model} is shown
in (e) for the constant flip angle scheme and (f) for varying flip angles.
}\label{fig:snr2} \end{figure}

\begin{figure}[h]
\centering
\begin{tabular}{ccc}
 \includegraphics[width=0.33\textwidth]{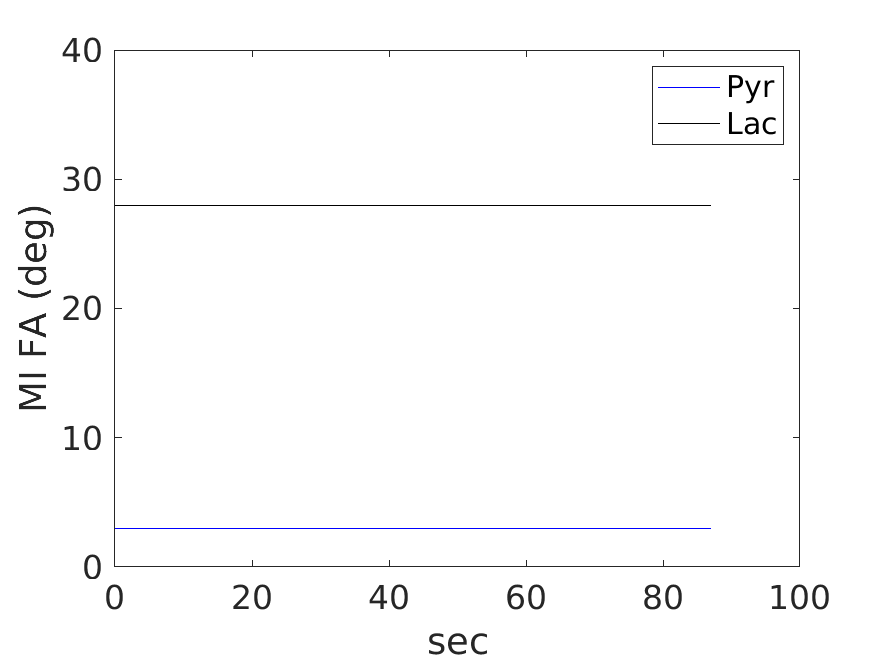}&
 \includegraphics[width=0.33\textwidth]{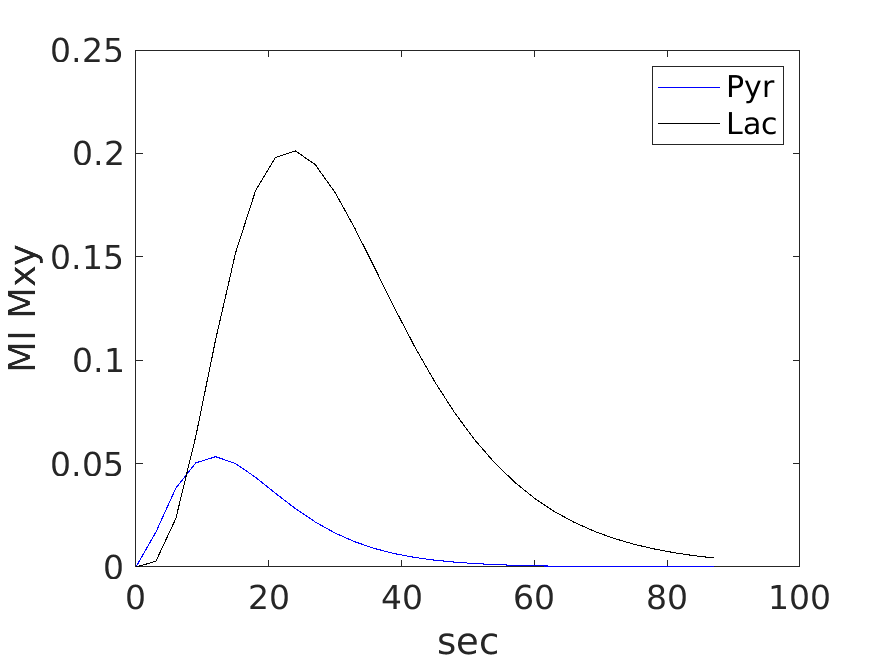}&
 \includegraphics[width=0.33\textwidth]{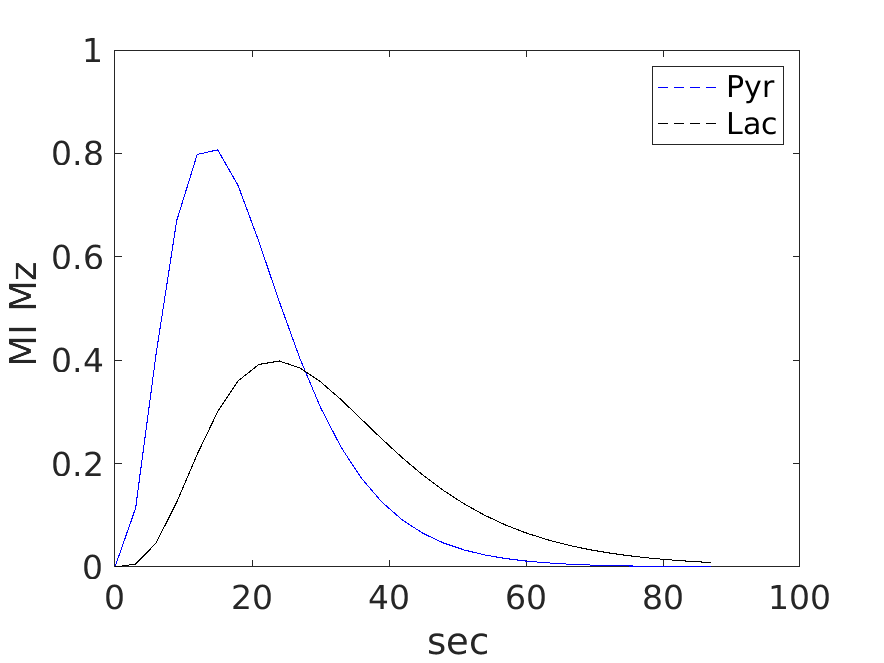}\\
(a) & (b) & (c) \\
 \includegraphics[width=0.33\textwidth]{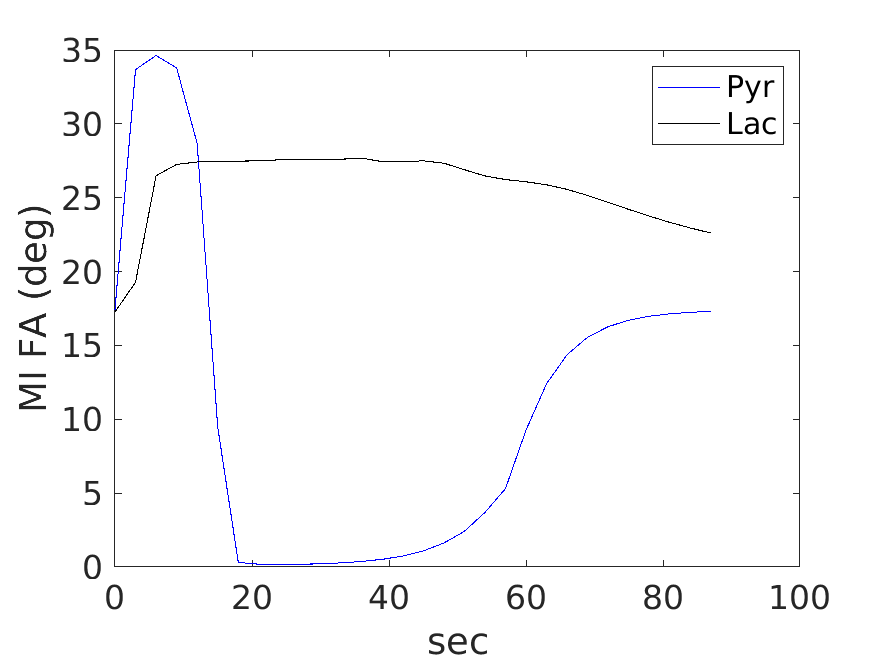}&
 \includegraphics[width=0.33\textwidth]{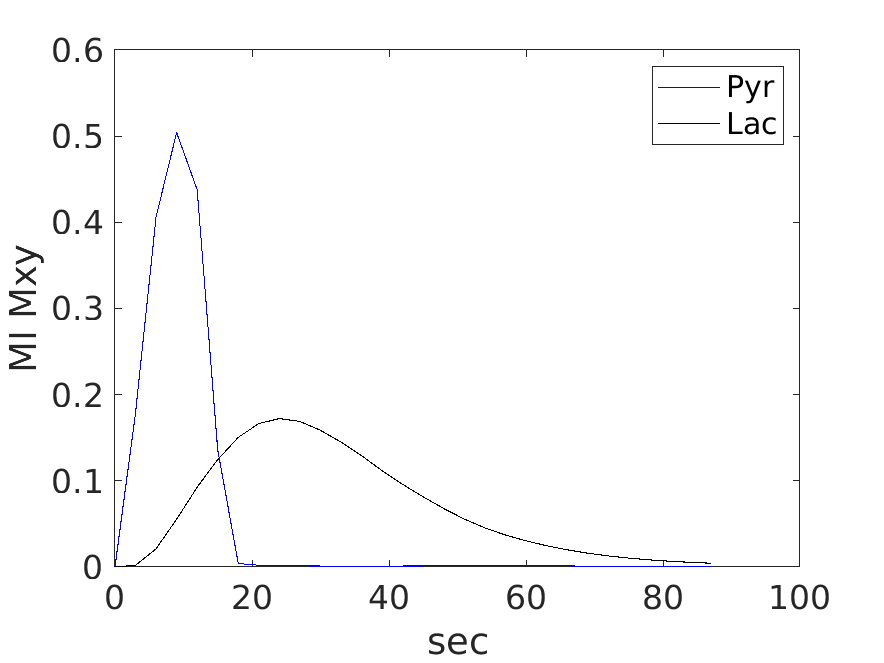}&
 \includegraphics[width=0.33\textwidth]{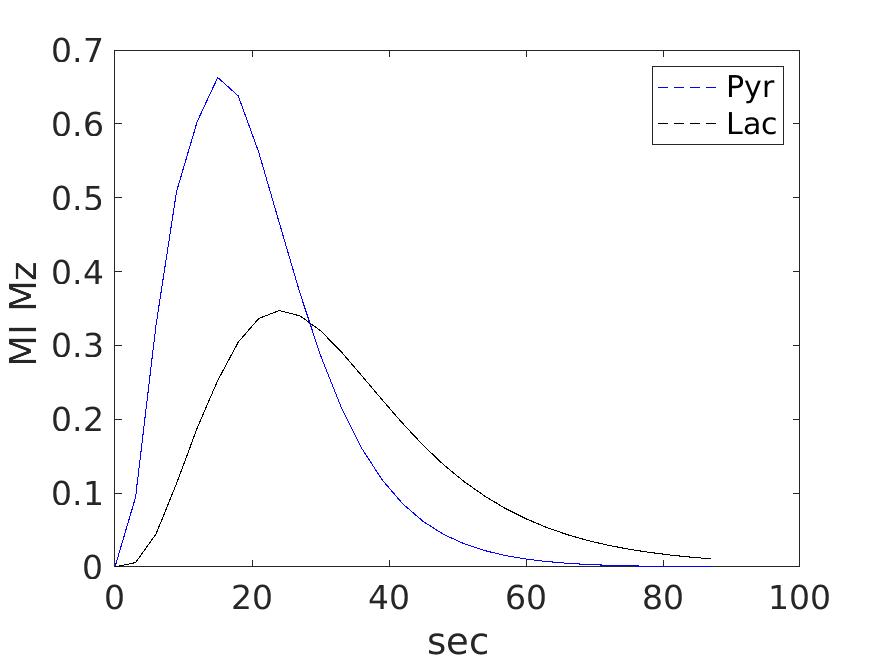}\\
(d) & (e) & (f) \\
\end{tabular}
\caption{Optimized design parameters along with the solution of the forward (LF)
model for noise $\sigma_z(20)$, i.e., $SNR = 20$. The optimized flip angle scheme is
shown for (a) constant flip angles throughout the acquisition (optimal angles are $\theta^k_P = 3$ degrees and $\theta^k_L = 28$ degrees, for all $1\leq k\leq N$) and (d)
allowing the flip angle to vary at each data acquisition.
The corresponding signal evolution of the transverse magnetization \eqn{eq:signal}
is shown in (b) for the constant flip angle scheme and (e) for
varying flip angles. Similarly, the longitudinal magnetization \eqn{eq:model} is shown
in (e) for the constant flip angle scheme and (f) for varying flip angles.
} \label{fig:snr20}
\end{figure}

As mentioned in \cref{s:mi}, multi-variate Gaussian is taken as a prior for uncertain model parameters, $\p = (k_{PL}, k_{ve}, \bar{t}_0)$. The mean and diagonal covariance matrix are fixed to 
$$
\color{black}
\mu_\p = (0.15, 0.05, 4), \qquad \Sigma_p = \mathrm{diag}(0.03, 0.01, 1.3) .
$$
For this  choice of mean and variance, all quadrature points 
for a fifth order Gauss-Hermite quadrature approximation of numerical integration were positive. 
The remaining model parameters are fixed according to Table~\ref{tab:modelParam}. 
Next, to fix the likelihood function, the Gaussian noise distribution, i.e., $\varepsilon \sim \calN(0, \sigma_z)$, needs to be fixed. 
To consider reasonable values of $\sigma_z$, first the reference peak pyruvate signal is calculated using the low-fidelity model with the default model and data parameters shown in Tables \ref{tab:modelParam} and \ref{tab:designParam}; it is found to be ${\sigp}_{ref} = 0.6173$. 
Then for different signal-to-noise ratios ($SNR$), the noise (standard deviation) in the individual signals, $\sigma_s$, and the standard deviation of the total signal, $\sigma_z$, are computed as follows:
\begin{equation}\label{eq:noiseSTD}
\sigma_s(SNR) = {\sigp}_{ref} / SNR, \qquad \sigma_z(SNR) = \sigma_s \sqrt{2N}, \qquad SNR \in \{2, 5, 10, 15, 20\},
\end{equation}
where $N$ is the total number of scans. 
For each $\sigma_z$ in the above list, optimal design parameters are obtained by solving the optimization problem \eqref{mimaxHz}. 
For simplicity, let $\kk_{OED_{\SNR}}$ denote the optimized design parameter corresponding 
to $\sigma_z(SNR)$. 
The rationale behind considering different $\SNR$ is that, in reality, data is expected to have 
varied signal-to-noise ratios and this is shown to impact the choice of optimal design parameters. 
The initial values of the design parameter $\kk$ are listed in
\cref{tab:designParam}. The number of scans, $N$, and repetition time, $TR_k$,
are fixed at the initial values. 
Figures \ref{fig:snr2} and \ref{fig:snr20} represent the optimal design parameters, $\kk_{OED_{\SNR}}$,
for the lower and upper bounds of SNR considered, $SNR = 2$ and $SNR=20$;
respectively. 
Results show that optimal design parameters vary depending on the $\SNR$ values;
however, results at  $SNR = 2$ and $SNR=20$ are representative of the
range of SNR considered. The optimal values of design parameters are shown in 
\cref{fig:snr2}(a), \cref{fig:snr2}(d), \cref{fig:snr20}(a) and \cref{fig:snr20}(d).
The time varying optimized design parameters are significantly different from the 
constant value flip angle scheme.
In \cref{fig:snr2} and \cref{fig:snr20}, HP pyruvate and lactate signals, $\sigp^k,\sigl^k$, $1\leq k \leq N$, are shown in (b) and (e), respectively. 
Moreover, in \cref{fig:snr2} and \cref{fig:snr20}, longitudinal magnetizations,  $\pyr^k, \lac^k$, are shown in (c) and (f), respectively. 

\subsection{Validation of the uncertainty reduction using optimal design parameters}\label{ss:OEDvalid}
The basic workflow in verification includes generating the samples of noisy data associated with different design parameters using either LF and HF model and then solving the inverse problem for model parameters  for each sample of data.  
From the recovered $k_{PL}$ for different samples of data, the mean and
the standard deviation is computed. Specifically, the standard deviation is used as a measure of the uncertainty in the recovered $k_{PL}$. 

Before discussing the results, steps in generating synthetic data are explained. 
Suppose $\kk_{OED_{\SNR}}$ is the optimal design parameter for the noise $\sigma_z(SNR)$ in the total signal, see \eqref{eq:noiseSTD}.  
Using either LF or HF models, ``ground truth" (signals at $N$ scans) is generated using $\kk_{OED_{\SNR}}$ design parameters ($TR_k$ is fixed following \cref{tab:designParam}) and default model parameters listed in \cref{tab:modelParam}. 
The data associated to $SNR$  in $\kk_{OED_{\SNR}}$ is denoted by $Y_{OED_{SNR}}$ and takes the form:
\[
Y_{OED_{SNR}} =
\bmatx{s^1_P & s^1_L \\ s^2_P & s^2_L \\ \cdot & \cdot \\ s^N_P & s^N_L}  .
\] 
In the case in which the LF model is used to generate ``ground truth", $s^i_P, s^i_L$ denote the pyruvate and lactate signals at \kth{i} scan, see \eqref{eq:signal}. 
If the data is generated from the HF model, these will be the average pyruvate and lactate signals within a voxel at scan time $t_i$, see \cref{sss:HFValid} for more details. 
A total of 25 samples of noisy data for different values of SNR are considered. Noisy samples are computed as follows:
\begin{equation}\label{eq:noisydata}
Y_{noisy, OED_{\SNR}, SNR_{data}} = f(Y_{OED_{\SNR}}, SNR_{data}) = Y_{OED_{SNR}} + \sigma_s(SNR_{data}) \bmatx{ a_1 & b_1 \\ a_2 & b_2 \\ \cdot & \cdot \\ a_N & b_N} ,
\end{equation}
where $\sigma_s(SNR_{data})$, given in \eqref{eq:noiseSTD}, is the amount of the noise that depends on the assumed SNR, $SNR_{data}$, $a_j, b_j\sim
\calN(0,1)$ for each $j=1,2,..,N$. 
Since the MI optimal solution is seen to vary with respect to SNR defined by the
total signal standard deviation, $\sigma_z$,
and SNR in the data, $SNR_{data}$, is expected to vary with pixelwise
location, 
a range of added noise values is considered to comprehensively evaluate the accuracy and precision of
the $k_{PL}$ parameter recovery. The noise values correspond to the previous SNR
range considered: $\sigma_s(SNR_{data}), SNR_{data} \in \{2, 5, 10, 15, 20\}$.
Validation results are now presented.

\subsubsection{Reduction in uncertainty in $k_{PL}$ for low-fidelity data}
\begin{figure}
\centering
\begin{tabular}{cccc} 
\includegraphics[width=.49\textwidth]{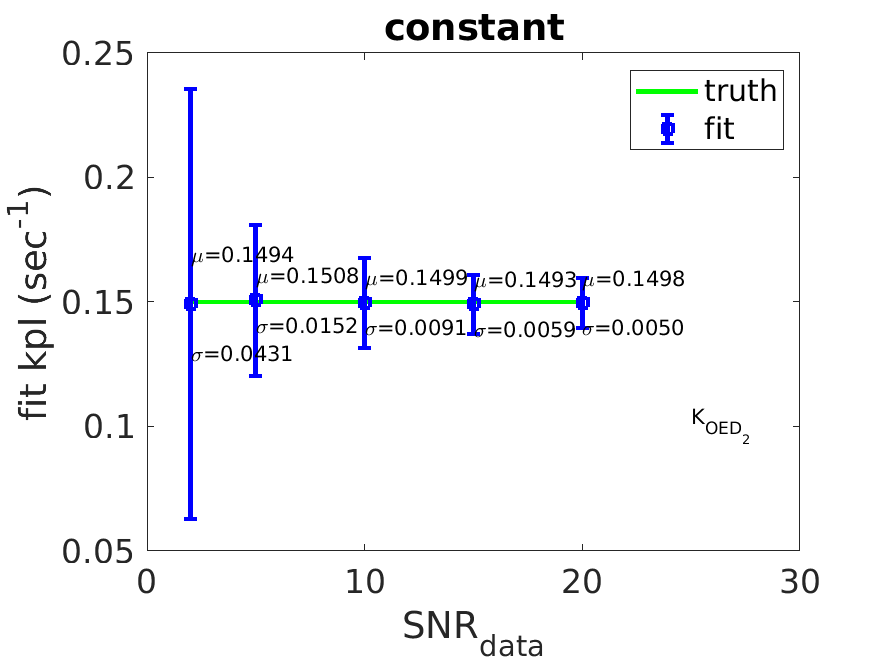} &
\includegraphics[width=.49\textwidth]{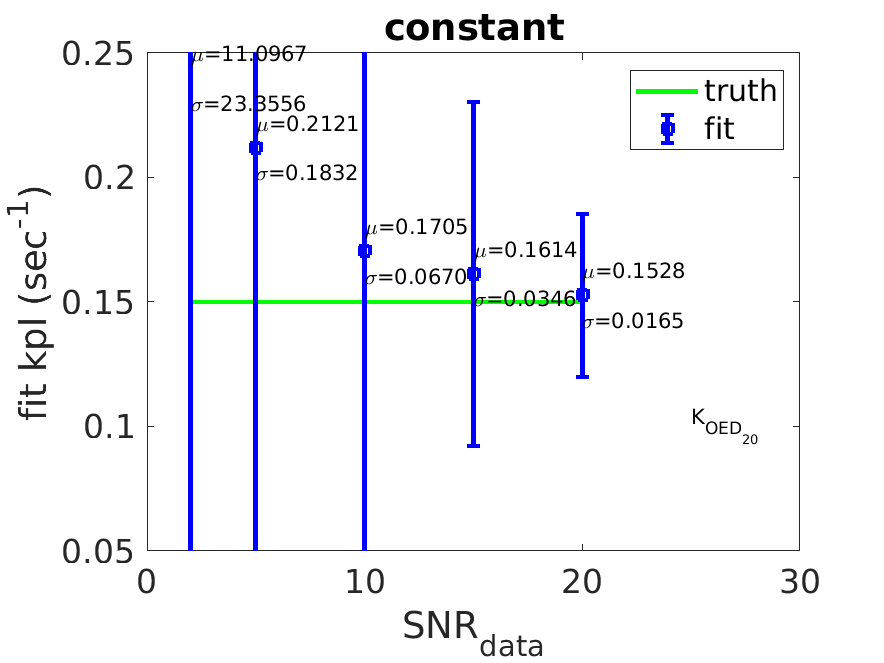} \\
(a) & (b) \\
\includegraphics[width=.49\textwidth]{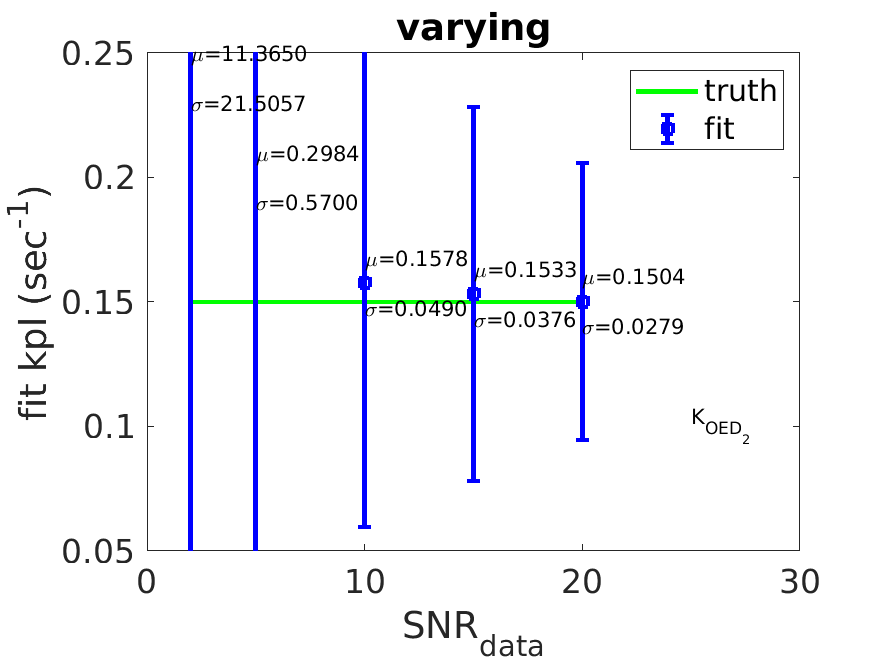} &
\includegraphics[width=.49\textwidth]{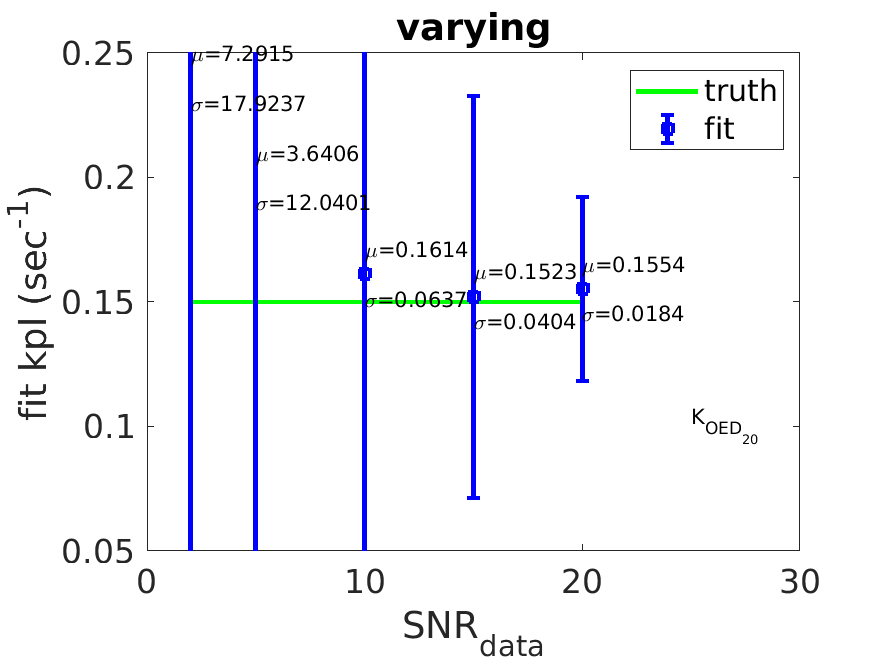}\\
(c) & (d) \\
\includegraphics[width=.49\textwidth]{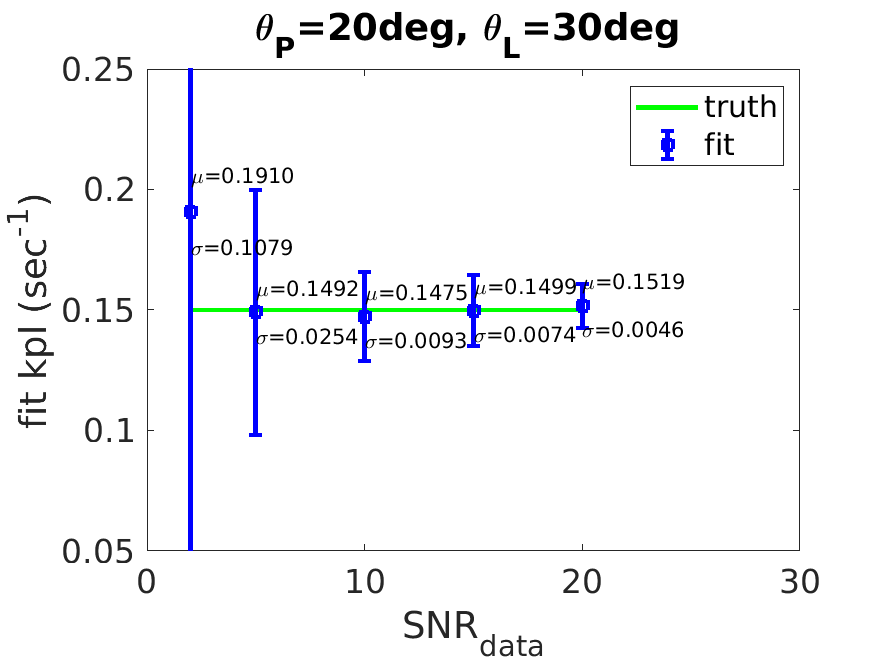}& \\
(e) &  \\
\end{tabular}
\caption{Plot of inferred $k_{PL}$ from 25 samples of noisy data obtained from LF model data.
The x-axis is the value of $\SNR_{data}$ corresponding to value of noise,
$\sigma_s(SNR_{data})$, added to the synthetic data for inference.
(a) and (c) correspond to 
$\kk_{OED_{2}}$ for a constant and varying flip angle scheme; respectively.
(b) and (d) correspond to 
$\kk_{OED_{20}}$ for a constant and varying flip angle scheme; respectively.
(e) provides
a control for the accuracy and precision
obtained for flip angles similar to values currently used in our human studies, $\theta^k_P=20$ and $\theta^k_L=30$.
}\label{fig:kplStatsLF}
\end{figure}

The LF model is used to generate data, $Y_{OED_{\SNR}}$ (for $\color{black}\SNR
= 2, 5, 10,  15, 20$), and uncertain parameters $\color{black}\p = (k_{PL}, k_{ve}, \bar{t}_0)$ in the LF model are recovered from the 25 samples of noisy data. 
The remaining model parameters are drawn from the \cref{tab:modelParam}.
{\color{black}In \cref{fig:kplStatsLF}, the statistics (mean and standard
deviation) corresponding to optimal design parameters shown in 
 \cref{fig:snr2} and \cref{fig:snr20} is presented.
The SNR (SNR to generate noisy data) along the x-axis corresponds to value of $\sigma_s(SNR_{data})$ added to the data.
The y-axis represents the mean and standard deviation of the $k_{PL}$ recovered from the inference.
The known value of $k_{PL}$ used to generate the noise corrupted data is shown for reference. 
Figures \ref{fig:kplStatsLF}(a) and \ref{fig:kplStatsLF}(c) correspond to 
$\kk_{OED_{2}}$ for a constant and varying flip angle scheme; respectively.
Figures \ref{fig:kplStatsLF}(b) and \ref{fig:kplStatsLF}(d) correspond to 
$\kk_{OED_{20}}$ for a constant and varying flip angle scheme; respectively.
Generally, an improvement in the accuracy and precision of the recovered parameter is seen
with increasing SNR.
The $k_{PL}$ recovered for  constant flip angles with 
$\kk_{OED_{2}}$ shows the best accuracy and precision for the noise corrupted
parameter recovery.
Figure \ref{fig:kplStatsLF}(e) provides a control for the accuracy and precision
obtained for flip angles similar to those currently in use in our
clinic~\cite{tang2021metabolic,NCT03830151}, $\theta^k_P=20$ and $\theta^k_L=30$.
Generally, results at
$\kk_{OED_{2}}$ are comparable to the current clinical pulse sequence
implementation. However, $\kk_{OED_{2}}$ demonstrates improved performance at low $SNR=2$.
}

\subsubsection{Recovery of rate parameter from the high-fidelity model-based data}\label{sss:HFValid}
A high-fidelity model is used to generate a somewhat more realistic spatially
varying data and further verify our results using an independent source of data. 
In this section, only the $k_{PL}$ parameter is treated as uncertain; all other model parameters were drawn from \cref{tab:modelParam}.

Consider a cubic domain $B$ containing the brain tissue domain 
$\Omega$ (discretization and simulation using HF model is discussed in \cref{s:discretization}). 
Let $B_h$ be a uniform discretization of $B$ with total $16^3$ cells. 
First, the HF model is simulated using the constant design parameter
$\kk_{OED_{SNR}}$ for $SNR = 2$, i.e., $\theta^k_P = 35$ degrees and $\theta^k_L = 28$ degrees, for all $1\leq k \leq N$, while all other parameters are fixed from \cref{tab:modelParam} and \cref{tab:designParam}. 
For the high-fidelity calculations, only constant $\kk_{OED_{2}}$ design parameter set was employed for the reasons that $\kk_{OED_2}$ is found to be more robust for large ranges of noises and the values in $\kk_{OED_2}$ are closer to what is currently being considered in clinical testing; see previous subsection. 
Next, the total pyruvate and lactate signals, $s^k_P, s^k_L$, $1\leq k\leq N$, are computed over each cell of $B_h$, i.e., $Y_{OED_{2}}$ at different cells of $B_h$ is obtained. 
Suppose $Y_{OED_{SNR}}^I$, $1\leq I\leq 16^3$, is the data for \kth{I} cell in grid $B_h$. 
Then, given the $SNR_{data} \in [2, 5, 10, 15, 20]$, the noisy data from $Y^I_{OED_2}$ for the \kth{I} cell is computed using
\begin{equation}\label{eq:noisydataHF}
Y^I_{noisy, OED_{\SNR}, SNR_{data}} = Y^I_{OED_{SNR}} + \sigma_s^{I}(SNR_{data}) \bmatx{ a_1 & b_1 \\ a_2 & b_2 \\ \cdot & \cdot \\ a_N & b_N} ,
\end{equation}
where  $\sigma^I_{s}(SNR_{data}) = \frac{s_{P_{ref}}^I}{SNR_{data}}$, $s^I_{P_{ref}}$ is the maximum pyruvate signal (maximum value in the first column of $Y_{OED_{SNR}}^I)$ for the \kth{I} cell, and $a_i, b_i, c_i, d_i$ are defined similar to \eqref{eq:noisydata}. 
Essentially, the noise model in this case varies from cell to cell and is scaled appropriately depending on the peak pyruvate signal for a given cell. 
The HP signals can be very small in many cells; peak signal in cells can be as small as $0.01$ based on simulation results.  
Therefore, adding noise $\sigma_s(SNR_{data}) = {\sigp}_{ref}/SNR_{data}$ based on one fixed reference signal ${\sigp}_{ref} = 0.6173$, see \eqref{eq:noiseSTD}, to data corresponding to cells with small peak pyruvate signal will produce meaningless results. 

\begin{table}[h!]
	\begin{center}
		\begin{tabular}{|c|c|c|c|} 
			\hline
			\textbf{Range} & \textbf{Number of cells} & \textbf{Range} & \textbf{Number of cells} \\
			\hline
			\hline
			$[1., 0.1)$ & $7$ & $[0.1, 0.01)$ & $12$\\
			$[0.01, 0.001)$ & $4$ & $[0.001, 0)$ & $2$\\
			\hline
		\end{tabular}
	\end{center}
	\caption{Four ranges of peak vascular pyruvate is considered to capture the variation in the uncertainty of parameters as cells move further from the vascular sources.}\label{tab:cellChoice}
\end{table}

Since it is expensive to solve an inverse problem on each cell in $B_h$ for all
ranges of $SNR_{data}$, 25 cells in $B_h$ are selected based on the peak value of HP
pyruvate in the vascular compartment to solve an inverse problem.
A total of four ranges of peak vascular pyruvate are considered and cells such that peak vascular
pyruvate of the cell, $\max_{k}\, \bar{\phi}_{PV}(t_k)$, is in the range is selected.
The range and number of selected cells are provided in \cref{tab:cellChoice}. 
Here, $\bar{\phi}_{PV}$ is the average of vascular pyruvate over a given cell.
Vascular pyruvate is nonzero only on $x\in \Omega_V$, where $\Omega_V$ is the vascular domain segmented from the MRI data, see ~\eqn{eq:vifvasc}.  
Thus the average vascular pyruvate will vary from cell to cell depending on the
magnitude of the the intersection between $\Omega_V$ and the cell.
The composition of cells based on the strength of vascular pyruvate agent is essential to capture the
variations in the uncertainty of recovered parameters both near and
far from the vascular sources.

\begin{figure}[h]
	\centering
	\includegraphics[width=0.65\textwidth]{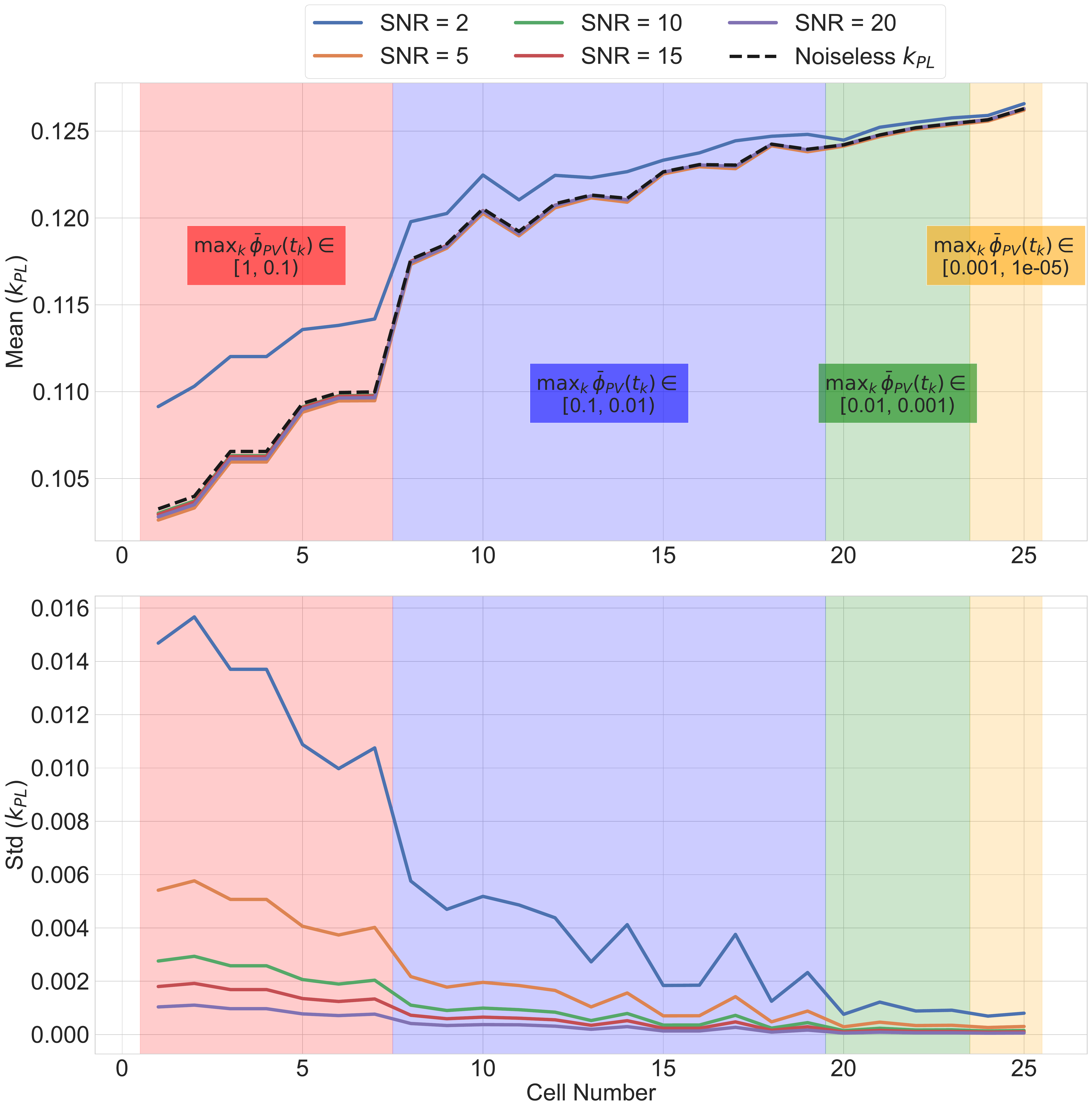}
	\caption{Plot of the mean and standard deviation of inferred $k_{PL}$
for 25 selected cells in $16^3$ grid. On x-axis, the cells are numbered and on
y-axis the value of the mean and standard deviation for specific cells are
plotted. Cells were selected based on the peak vascular pyruvate,
$\max_{k}\, \bar{\phi}_{PV}(t_k)$, where $\bar{\phi}_{PV}$ is the average of vascular
pyruvate over a given cell, belonging to four range of values as shown in the
plot. Color of each curve is related to the $SNR_{data}$ used in computing noisy
data.
 }\label{fig:kplStats}
\end{figure}

In \cref{fig:kplStats}, the mean and standard deviation of inferred $k_{PL}$ for different cells and different $\SNR_{data}$ values are shown. Here, only one design parameter set, $\kk_{OED_{2}}$, is considered. 
The X-axis in \cref{fig:kplStats} identifies the cell number (total 25 cells are considered) while in the Y-axis the mean (on top plot) and standard deviation (on bottom plot) of recovered $k_{PL}$ are displayed. 
Different colors signify different values of $SNR_{data}$. 
`Noiseless $k_{PL}$' is the one that is obtained by solving an inverse problem on data from specific cells without adding noise. 

From \cref{fig:kplStats}, the following key observations can be made:
\begin{itemize}
	\item \textbf{Decrease in the uncertainty with decreasing noise in the
data}: The smaller $\sigma_s(SNR_{data})$ results in smaller overall uncertainty
in the recovered $k_{PL}$. Further, the recovered $k_{PL}$ from noisy data are
seen to converge to the noiseless $k_{PL}$ obtained without the addition of any noise.

	\item \textbf{Uncertainty is stronger for cells with high peak vascular pyruvate}: In cells with stronger vascular 
pyruvate, the uncertainty in recovered $k_{PL}$ is higher. 
In fact as the peak vascular pyruvate in cells decrease (cell number low to high), the recovered $k_{PL}$, 
e.g., mean $k_{PL}$ in the \cref{fig:kplStats}, gets closer to the $k_{PL} = 0.15$ employed in the high-fidelity simulation. 
Therefore, the results suggest that in the cells with high impact from a vascular source, the inconsistency between the high-fidelity model and low-fidelity is higher. 
\end{itemize}

\section{Discussion}\label{s:disc}
Within the mutual information based optimal experimental design formulation, a time varying
flip angle scheme is seen to provide 
significant differences in the flip angles 
as compared to the case when excitation angles are fixed to a constant value.
Indeed, the varying flip angle scheme leads to a higher parameter optimization
that is able to further improve the quantitative value of mutual information 
over the constant flip angle scheme. However, as seen in \cref{fig:kplStatsLF},
the constant flip angle scheme leads to the best accuracy and precision when
considering the inference from noise-corrupted data. The time varying
flip angle scheme is seen to be most sensitive to noise corruption of the expected
signal and is generally seen to have the highest variance and bias in the
parameter recovery.

The reduction in the recovered
variance is seen to be correlated with the assumed noise value added to the
data. Intuitively, less noise resulted in
less variance in the parameter recovery. Less intuitively, the 
optimal MI solutions for 
flip angles 
are seen to vary with the noise value of the 
signal conditional probability model.
The greatest reduction in measurement uncertainty is seen for the MI optimal solution corresponding to
low SNR.
The excitation angles optimized for the higher SNR condition drop the pyruvate
excitation angle to zero after 20 seconds while simultaneously increasing the
lactate flip angle. This is in contrast to the low SNR case where both pyruvate
and lactate flip angles maintain an continual signal. This could
be due to the system being so signal limited for the low SNR case that it is
forced to leverage the pyruvate signal to extract any additional information
about the metabolic exchange rate. As opposed to the high SNR case where the
abundant lactate signal is sufficient to accurately determine the metabolic
exchange rate and measuring pyruvate signal after the initial bolus is therefore unnecessarily.

This work also highlights some issues observed when using the high-fidelity spatially varying model generated ``ground truth" in recovering the model parameters in the low-fidelity spatially invariant model. 
The recovered parameters of the low-fidelity model are not same as the values considered in generating ``ground truth"; further this difference is higher at locations with stronger vascular pyruvate source. 
Thus, this work shows that it is possible that the difference between $k_{PL}$ in real tissue and the $k_{PL}$ values recovered using the low-fidelity model could be large. 

This work considers uncertainty in the
vascular-tissue exchange parameter, bolus arrival time, and rate constants. However, a more comprehensive evaluation 
of additional uncertain parameters would be to additionally consider 
the stability of our results.
The numerical computation in this work is also limited by the quadrature
scheme for numerical integration of the mutual information integrals.
Adding additional sources of uncertainty suffers from the well-known curse of
dimensionality~\cite{berchtold1998pyramid} and alternative integration schemes
such as Markov chain Monte Carlo may be more effective. 

Further, the current approach considers the real component of the readout and assumes SNR such that Gaussian statistics is an
appropriate  noise model for the signal acquisition. 
Rician statistics~\cite{gudbjartsson1995rician} is known to be more appropriate
as the noise model for low SNR and
the low SNR range is expected to be
more important toward the end of the HP data acquisition as the signal decays.
Rician statistics will be considered in future efforts to
optimize acquisition parameters at low SNR or when considering both the real and
imaginary components of the signal magnitude.

Alternative high-fidelity models may also be considered to determine optimal
design parameters and to recover model parameters from the data. However, for
such an approach to work, a realistic high-fidelity model is needed keeping in
mind the major factors in HP-MRI physics.  This work presents a high-fidelity
model based on various simplifying assumptions as a first step. 
Moreover, the model in this work was limited to just two constituents. 
Additional model fidelity may include permutations of lactate and pyruvate
that are endogenous as well as hyperpolarized. 
Intravascular, extracellular, and intracellular species may also be considered. 
Additional formulations may also consider the impact of blood flow in the 
simulations directly though Dirichlet boundary conditions, as convective
transport through porous media~\cite{fuentes2020imaging}, or as a sophisticated
3D-1D coupling with vasculature treated as 1D curvilinear segments\cite{fritz2020analysis, fritz2021modeling}.

In summary, our results suggest that the constant flip angle scheme corresponding to
$\kk_{OED_{2}}$ is the best choice in terms of accuracy and precision of the  parameter recovery.
Results at $\kk_{OED_{2}}$, $\theta^k_P=35$ and $\theta^k_L=28$, are comparable to the current clinical pulse sequence
implementations, $\theta^k_P=20$ and $\theta^k_L=30$, and demonstrate an improved performance at low $SNR_{data}=2$.
Further, the constant flip angle scheme may represent a
practical choice for
implementation on the pulse sequence hardware. 

\section*{Acknowledgments}
We acknowledge Reshmi Patel and Collin J. Harlan for sharing the insights from their work. The work of PKJ and JTO was supported by the U.S. Department of Energy, Office of Science, USA, Office of Advanced Scientific Computing Research, Mathematical Multifaceted Integrated Capability Centers (MMICCS), under Award Number DE-SC0019303.
DF, JB, and DS are supported under R21CA249373A1. 
This work was supported as one of the pilot projects by the MDACC-Oden Institute-TACC joint initiative. JB, DF, and PKJ acknowledge this support.


\begin{thebibliography}{10}

\bibitem{nelson2013metabolic}
Sarah~J Nelson, John Kurhanewicz, Daniel~B Vigneron, Peder~EZ Larson, Andrea~L
  Harzstark, Marcus Ferrone, Mark Van~Criekinge, Jose~W Chang, Robert Bok,
  Ilwoo Park, et~al.
\newblock Metabolic imaging of patients with prostate cancer using
  hyperpolarized [1-13c] pyruvate.
\newblock {\em Science translational medicine}, 5(198):198ra108--198ra108,
  2013.

\bibitem{Bankson2015}
J.A. Bankson, C.M. Walker, M.S. Ramirez, W.~Stefan, D.~Fuentes, M.E. Merritt,
  J.~Lee, V.C. Sandulache, Y.~Chen, L.~Phan, P.-C. Chou, A.~Rao, S.-C.J. Yeung,
  M.-H. Lee, D.~Schellingerhout, C.A. Conrad, C.~Malloy, A.D. Sherry, S.Y. Lai,
  and J.D. Hazle.
\newblock {Kinetic modeling and constrained reconstruction of hyperpolarized
  [1-{\textless}sup{\textgreater}13{\textless}/sup{\textgreater}C]-pyruvate
  offers improved metabolic imaging of tumors}.
\newblock {\em Cancer Research}, 75(22), 2015.

\bibitem{kurhanewicz2019hyperpolarized}
John Kurhanewicz, Daniel~B Vigneron, Jan~Henrik Ardenkjaer-Larsen, James~A
  Bankson, Kevin Brindle, Charles~H Cunningham, Ferdia~A Gallagher, Kayvan~R
  Keshari, Andreas Kjaer, Christoffer Laustsen, et~al.
\newblock Hyperpolarized 13c mri: path to clinical translation in oncology.
\newblock {\em Neoplasia}, 21(1):1--16, 2019.

\bibitem{granlund2020hyperpolarized}
Kristin~L Granlund, Sui-Seng Tee, Hebert~A Vargas, Serge~K Lyashchenko,
  Ed~Reznik, Samson Fine, Vincent Laudone, James~A Eastham, Karim~A Touijer,
  Victor~E Reuter, et~al.
\newblock Hyperpolarized mri of human prostate cancer reveals increased lactate
  with tumor grade driven by monocarboxylate transporter 1.
\newblock {\em Cell metabolism}, 31(1):105--114, 2020.

\bibitem{miloushev2016hyperpolarization}
Vesselin~Z Miloushev, Kayvan~R Keshari, and Andrei~I Holodny.
\newblock Hyperpolarization mri: preclinical models and potential applications
  in neuroradiology.
\newblock {\em Topics in magnetic resonance imaging: TMRI}, 25(1):31, 2016.

\bibitem{miloushev2018metabolic}
Vesselin~Z Miloushev, Kristin~L Granlund, Rostislav Boltyanskiy, Serge~K
  Lyashchenko, Lisa~M DeAngelis, Ingo~K Mellinghoff, Cameron~W Brennan, Vivian
  Tabar, T~Jonathan Yang, Andrei~I Holodny, et~al.
\newblock Metabolic imaging of the human brain with hyperpolarized 13c pyruvate
  demonstrates 13c lactate production in brain tumor patients.
\newblock {\em Cancer research}, 78(14):3755--3760, 2018.

\bibitem{aggarwal2017hyperpolarized}
Rahul Aggarwal, Daniel~B Vigneron, and John Kurhanewicz.
\newblock Hyperpolarized 1-[13c]-pyruvate magnetic resonance imaging detects an
  early metabolic response to androgen ablation therapy in prostate cancer.
\newblock {\em European urology}, 72(6):1028, 2017.

\bibitem{gallagher2020imaging}
Ferdia~A Gallagher, Ramona Woitek, Mary~A McLean, Andrew~B Gill, Raquel~Manzano
  Garcia, Elena Provenzano, Frank Riemer, Joshua Kaggie, Anita Chhabra, Stephan
  Ursprung, et~al.
\newblock Imaging breast cancer using hyperpolarized carbon-13 mri.
\newblock {\em Proceedings of the National Academy of Sciences},
  117(4):2092--2098, 2020.

\bibitem{woitek2020hyperpolarized}
Ramona Woitek, Mary~A McLean, Andrew~B Gill, James~T Grist, Elena Provenzano,
  Andrew~J Patterson, Stephan Ursprung, Turid Torheim, Fulvio Zaccagna, Matthew
  Locke, et~al.
\newblock Hyperpolarized 13c mri of tumor metabolism demonstrates early
  metabolic response to neoadjuvant chemotherapy in breast cancer.
\newblock {\em Radiology: Imaging Cancer}, 2(4):e200017, 2020.

\bibitem{warburg1956origin}
Otto Warburg.
\newblock On the origin of cancer cells.
\newblock {\em Science}, 123(3191):309--314, 1956.

\bibitem{vander2009understanding}
Matthew~G Vander~Heiden, Lewis~C Cantley, and Craig~B Thompson.
\newblock Understanding the warburg effect: the metabolic requirements of cell
  proliferation.
\newblock {\em science}, 324(5930):1029--1033, 2009.

\bibitem{grist2019quantifying}
James~T Grist, Mary~A McLean, Frank Riemer, Rolf~F Schulte, Surrin~S Deen,
  Fulvio Zaccagna, Ramona Woitek, Charlie~J Daniels, Joshua~D Kaggie, Tomasz
  Matys, et~al.
\newblock Quantifying normal human brain metabolism using hyperpolarized
  [1--13c] pyruvate and magnetic resonance imaging.
\newblock {\em NeuroImage}, 189:171--179, 2019.

\bibitem{lee2020lactate}
Casey~Y Lee, Hany Soliman, Benjamin~J Geraghty, Albert~P Chen, Kim~A Connelly,
  Ruby Endre, William~J Perks, Chris Heyn, Sandra~E Black, and Charles~H
  Cunningham.
\newblock Lactate topography of the human brain using hyperpolarized 13c-mri.
\newblock {\em NeuroImage}, 204:116202, 2020.

\bibitem{autry2019comparison}
Adam~W Autry, Jeremy~W Gordon, Lucas Carvajal, Azma Mareyam, Hsin-Yu Chen,
  Ilwoo Park, Daniele Mammoli, Maryam Vareth, Susan~M Chang, Lawrence~L Wald,
  et~al.
\newblock Comparison between 8-and 32-channel phased-array receive coils for in
  vivo hyperpolarized 13c imaging of the human brain.
\newblock {\em Magnetic resonance in medicine}, 82(2):833--841, 2019.

\bibitem{walker2019effects}
Christopher~M Walker, David Fuentes, Peder~EZ Larson, Vikas Kundra, Daniel~B
  Vigneron, and James~A Bankson.
\newblock Effects of excitation angle strategy on quantitative analysis of
  hyperpolarized pyruvate.
\newblock {\em Magnetic resonance in medicine}, 81(6):3754--3762, 2019.

\bibitem{maidens2015optimal}
John Maidens, Peder~EZ Larson, and Murat Arcak.
\newblock Optimal experiment design for physiological parameter estimation
  using hyperpolarized carbon-13 magnetic resonance imaging.
\newblock In {\em 2015 American Control Conference (ACC)}, pages 5770--5775.
  IEEE, 2015.

\bibitem{maidens2016optimizing}
John Maidens, Jeremy~W Gordon, Murat Arcak, and Peder~EZ Larson.
\newblock Optimizing flip angles for metabolic rate estimation in
  hyperpolarized carbon-13 mri.
\newblock {\em IEEE transactions on medical imaging}, 35(11):2403--2412, 2016.

\bibitem{marseille1996bayesian}
GJ~Marseille, R~De~Beer, M~Fuderer, AF~Mehlkopf, and D~van Ormondt.
\newblock Bayesian estimation of mr images from incomplete raw data.
\newblock In {\em Maximum Entropy and Bayesian Methods}, pages 13--22.
  Springer, 1996.

\bibitem{brihuega2003optimization}
Oscar Brihuega-Moreno, Frank~P Heese, and Laurance~D Hall.
\newblock Optimization of diffusion measurements using cramer-rao lower bound
  theory and its application to articular cartilage.
\newblock {\em Magnetic resonance in medicine}, 50(5):1069--1076, 2003.

\bibitem{poot2010optimal}
Dirk~HJ Poot, Arnold~J den Dekker, Eric Achten, Marleen Verhoye, and Jan
  Sijbers.
\newblock Optimal experimental design for diffusion kurtosis imaging.
\newblock {\em Medical Imaging, IEEE Transactions on}, 29(3):819--829, 2010.

\bibitem{cercignani2006optimal}
Mara Cercignani and Daniel~C Alexander.
\newblock Optimal acquisition schemes for in vivo quantitative magnetization
  transfer mri.
\newblock {\em Magnetic resonance in medicine}, 56(4):803--810, 2006.

\bibitem{reeves1999sequential}
Stanley~J Reeves and Zhao Zhe.
\newblock Sequential algorithms for observation selection.
\newblock {\em Signal Processing, IEEE Transactions on}, 47(1):123--132, 1999.

\bibitem{ji2008bayesian}
Shihao Ji, Ya~Xue, and Lawrence Carin.
\newblock Bayesian compressive sensing.
\newblock {\em Signal Processing, IEEE Transactions on}, 56(6):2346--2356,
  2008.

\bibitem{seeger2010optimization}
Matthias Seeger, Hannes Nickisch, Rolf Pohmann, and Bernhard Sch{\"o}lkopf.
\newblock Optimization of k-space trajectories for compressed sensing by
  bayesian experimental design.
\newblock {\em Magnetic resonance in medicine}, 63(1):116--126, 2010.

\bibitem{Cover2012}
Thomas~M Cover and Joy~A Thomas.
\newblock {\em Elements of information theory}.
\newblock John Wiley \& Sons, 2012.

\bibitem{oden2010general}
J~Tinsley Oden, Andrea Hawkins, and Serge Prudhomme.
\newblock General diffuse-interface theories and an approach to predictive
  tumor growth modeling.
\newblock {\em Mathematical Models and Methods in Applied Sciences},
  20(03):477--517, 2010.

\bibitem{riviere2008discontinuous}
Beatrice Riviere.
\newblock {\em Discontinuous Galerkin methods for solving elliptic and
  parabolic equations: theory and implementation}.
\newblock SIAM, 2008.

\bibitem{peacemanfundamentals}
Donald~W Peaceman.
\newblock {\em Fundamentals of numerical reservoir simulation}.
\newblock Elsevier, 1977.

\bibitem{faust1979geothermal}
Charles~R Faust and James~W Mercer.
\newblock Geothermal reservoir simulation: 1. mathematical models for
  liquid-and vapor-dominated hydrothermal systems.
\newblock {\em Water resources research}, 15(1):23--30, 1979.

\bibitem{Gerstner1998}
Thomas Gerstner and Michael Griebel.
\newblock {Numerical integration using sparse grids}.
\newblock {\em Numerical Algorithms}, 18:209--232, 1998.

\bibitem{Niederreiter1992}
H.~Niederreiter.
\newblock {\em {Random Number Generation and Monte Carlo Methods}}.
\newblock 1992.

\bibitem{robert2013monte}
Christian Robert and George Casella.
\newblock {\em Monte Carlo statistical methods}.
\newblock Springer Science \& Business Media, 2013.

\bibitem{gordon1996computer}
Sanford Gordon and Bonnie~J McBride.
\newblock {\em Computer program for calculation of complex chemical equilibrium
  compositions and applications}.
\newblock Citeseer, 1996.

\bibitem{Sloan1994}
I~H Sloan, S~Joe, and S.L.M.S. Joe.
\newblock {\em {Lattice Methods for Multiple Integration}}.
\newblock Oxford science publications. Clarendon Press, 1994.

\bibitem{Genz1987}
Alan Genz.
\newblock {\em {A Package for Testing Multiple Integration Subroutines}}, pages
  337--340.
\newblock Springer Netherlands, Dordrecht, 1987.

\bibitem{VanDooren1976}
Paul van Dooren and Luc de~Ridder.
\newblock {An adaptive algorithm for numerical integration over an
  n-dimensional cube}.
\newblock {\em Journal of Computational and Applied Mathematics},
  2(3):207--217, 1976.

\bibitem{Barron1994}
Andrew~R. Barron.
\newblock {Approximation and Estimation Bounds for Artificial Neural Networks}.
\newblock {\em Machine Learning}, 14(1):115--133, 1994.

\bibitem{Cavagnaro2010}
DR~Daniel~R Cavagnaro, JI~Jay~I Myung, Mark~a Pitt, and Janne V~JV Kujala.
\newblock {Adaptive design optimization: A mutual information-based approach to
  model discrimination in cognitive science}.
\newblock {\em Neural computation}, 22(1956):1--15, 2010.

\bibitem{Drovandi2014}
Christopher~C. Drovandi, James~M. McGree, and Anthony~N. Pettitt.
\newblock {A sequential Monte Carlo algorithm to incorporate model uncertainty
  in Bayesian sequential design}.
\newblock {\em Journal of Computational and Graphical Statistics}, 23(1):3--24,
  2014.

\bibitem{Ryan2014}
Elizabeth~G. Ryan, Christopher~C. Drovandi, M.~Helen Thompson, and Anthony~N.
  Pettitt.
\newblock {Towards Bayesian experimental design for nonlinear models that
  require a large number of sampling times}.
\newblock {\em Computational Statistics and Data Analysis}, 70:45--60, 2014.

\bibitem{Ryan2003}
Kenneth~J Ryan.
\newblock {Estimating Expected Information Gains for Experimental Designs With
  Application to the Random Fatigue-Limit Model}.
\newblock {\em Journal of Computational and Graphical Statistics},
  12(3):585--603, 2003.

\bibitem{Ryan2016}
Elizabeth~G. Ryan, Christopher~C. Drovandi, James~M. Mcgree, and Anthony~N.
  Pettitt.
\newblock {A Review of Modern Computational Algorithms for Bayesian Optimal
  Design}.
\newblock {\em International Statistical Review}, 84(1):128--154, 2016.

\bibitem{mitchell2020information}
Drew~P Mitchell, Ken-Pin Hwang, James~A Bankson, R~Jason Stafford, Suchandrima
  Banerjee, Naoyuki Takei, and David Fuentes.
\newblock An information theory model for optimizing quantitative magnetic
  resonance imaging acquisitions.
\newblock {\em Physics in Medicine \& Biology}, 65(22):225008, 2020.

\bibitem{tang2021metabolic}
Shuyu Tang, Maxwell~V Meng, James~B Slater, Jeremy~W Gordon, Daniel~B Vigneron,
  Bradley~A Stohr, Peder~EZ Larson, and Zhen~Jane Wang.
\newblock Metabolic imaging with hyperpolarized 13c pyruvate magnetic resonance
  imaging in patients with renal tumorsinitial experience.
\newblock {\em Cancer}, 127(15):2693--2704, 2021.

\bibitem{NCT03830151}
D.~Schellingerhout.
\newblock Hyperpolarized carbon c 13 pyruvate in diagnosing glioma in patients
  with brain tumors, July 2022.

\bibitem{berchtold1998pyramid}
Stefan Berchtold, Christian B{\"o}hm, and Hans-Peter Kriegal.
\newblock The pyramid-technique: Towards breaking the curse of dimensionality.
\newblock In {\em Proceedings of the 1998 ACM SIGMOD international conference
  on Management of data}, pages 142--153, 1998.

\bibitem{gudbjartsson1995rician}
H{\'a}kon Gudbjartsson and Samuel Patz.
\newblock The rician distribution of noisy mri data.
\newblock {\em Magnetic resonance in medicine}, 34(6):910--914, 1995.

\bibitem{fuentes2020imaging}
D~Fuentes, E~Thompson, M~Jacobsen, A~Colleen Crouch, RR~Layman, B~Riviere, and
  E~Cressman.
\newblock Imaging-based characterization of convective tissue properties.
\newblock {\em International Journal of Hyperthermia}, 37(3):155--163, 2020.

\bibitem{fritz2020analysis}
Marvin Fritz, Prashant~K. Jha, Tobias Köppl, J.~Tinsley Oden, and Barbara
  Wohlmuth.
\newblock Analysis of a new multispecies tumor growth model coupling 3d
  phase-fields with a 1d vascular network.
\newblock {\em Nonlinear Analysis: Real World Applications}, 61:103331, 2021.

\bibitem{fritz2021modeling}
Marvin Fritz, Prashant~K Jha, Tobias K{\"o}ppl, J~Tinsley Oden, Andreas Wagner,
  and Barbara Wohlmuth.
\newblock Modeling and simulation of vascular tumors embedded in evolving
  capillary networks.
\newblock {\em Computer Methods in Applied Mechanics and Engineering},
  384:113975, 2021.

\bibitem{menze2014multimodal}
Bjoern~H Menze, Andras Jakab, Stefan Bauer, Jayashree Kalpathy-Cramer, Keyvan
  Farahani, Justin Kirby, Yuliya Burren, Nicole Porz, Johannes Slotboom, Roland
  Wiest, et~al.
\newblock The multimodal brain tumor image segmentation benchmark (brats).
\newblock {\em IEEE transactions on medical imaging}, 34(10):1993--2024, 2014.

\bibitem{avants2011open}
Brian~B Avants, Nicholas~J Tustison, Jue Wu, Philip~A Cook, and James~C Gee.
\newblock An open source multivariate framework for n-tissue segmentation with
  evaluation on public data.
\newblock {\em Neuroinformatics}, 9(4):381--400, 2011.

\bibitem{frangi1998multiscale}
Alejandro~F Frangi, Wiro~J Niessen, Koen~L Vincken, and Max~A Viergever.
\newblock Multiscale vessel enhancement filtering.
\newblock In {\em International conference on medical image computing and
  computer-assisted intervention}, pages 130--137. Springer, 1998.

\bibitem{otsu1979threshold}
Nobuyuki Otsu.
\newblock A threshold selection method from gray-level histograms.
\newblock {\em IEEE transactions on systems, man, and cybernetics},
  9(1):62--66, 1979.

\bibitem{AlnaesBlechta2015a}
Martin~S. Aln{\ae}s, Jan Blechta, Johan Hake, August Johansson, Benjamin
  Kehlet, Anders Logg, Chris Richardson, Johannes Ring, Marie~E. Rognes, and
  Garth~N. Wells.
\newblock The fenics project version 1.5.
\newblock {\em Archive of Numerical Software}, 3(100), 2015.

\bibitem{LoggMardalEtAl2012a}
Anders Logg, Kent-Andre Mardal, Garth~N. Wells, et~al.
\newblock {\em Automated Solution of Differential Equations by the Finite
  Element Method}.
\newblock Springer, 2012.

\end{thebibliography}


\appendix

\section{Development of a high-fidelity model for HP-MRI}\label{s:derivationPDEModel}
\begin{figure}
\centering
\includegraphics[width=0.5\textwidth]{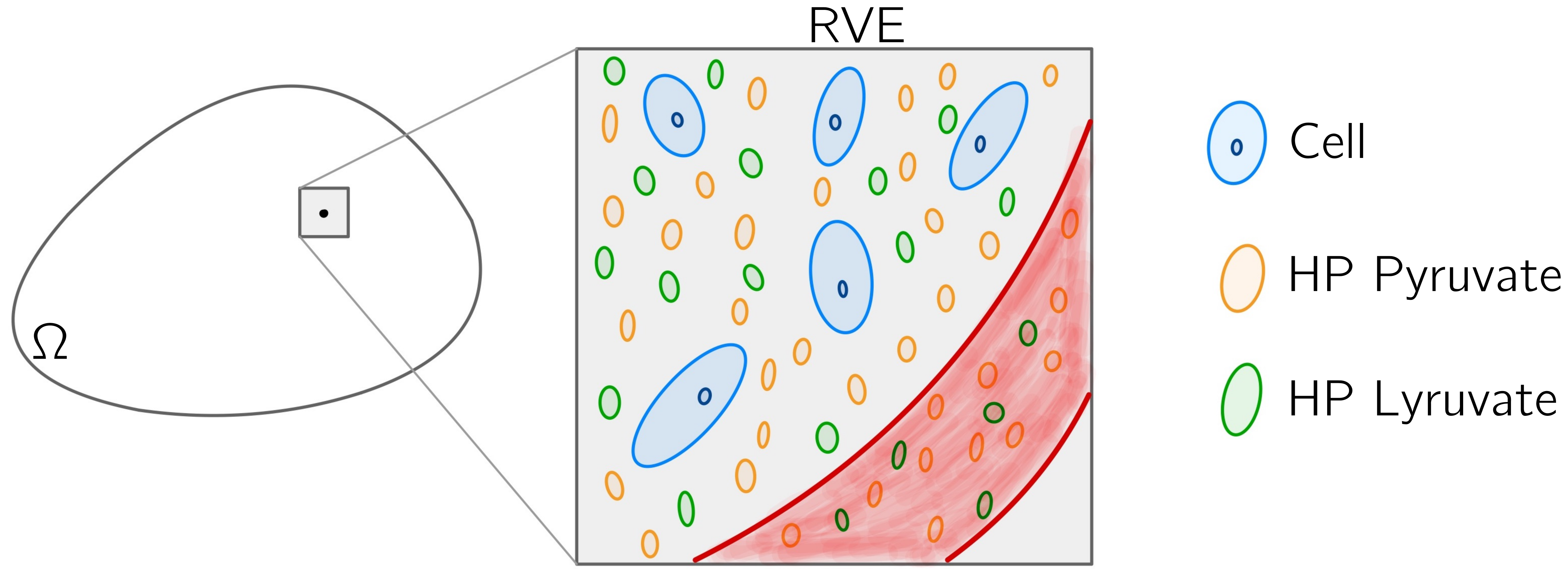}
\caption{Schematics of an RVE at sub-tissue scale with basic constituents and compartments.}\label{fig:rve}
\end{figure}

Let $\Omega \subset \bbR^3$ represent a three-dimensional tissue domain and let $\Omega_V \subset \Omega$ is the part of the tissue occupied by the vascular structure. 
Consider an RVE (Representative Volume Element) of sub-tissue length scale ($\approx$ 1 mm) at some point $\bx \in \Omega$. 
A simple depiction of RVE 
is shown in \cref{fig:rve} which highlights three key compartments, namely, interstitial, vascular, and cellular. 
Suppose $V$ is the total volume of RVE, and $V_i$, $V_v$, and $V_c$ are the volumes of interstitial, vascular, and cellular compartments, respectively, such that $V = V_i + V_v + V_c$. 
Further, let $\sigma_a = V_a / V$, $a\in \{i, v, c\}$, give the volume fractions of compartments. 

While each compartment may consist of variety of constituents, HP pyruvate and lactate in the interstitial and vascular compartments are of particular interest. 
Let $\phi_P = V_P / V$ and $\phi_L = V_L/ V$ denote the volume fraction of HP pyruvate and lactate in interstitium ($V_P, V_L$ are the volumes of HP agents in the interstitial compartment of an RVE) 
while $\phi_{\sim PL}$ giving the volume fraction of all other constituents in the interstitial compartment. 
Volume fractions in vascular compartment, $\phi_{PV}, \phi_{LV}, \phi_{\sim PLV}$, and in cellular compartment, $\phi_{PC}, \phi_{LC}, \phi_{\sim PLC}$, are defined in a similar fashion. 
For a generic RVE, following saturation condition must hold, i.e.,
\begin{equation*}
\phi_P + \phi_L + \phi_{\sim PL} = \sigma_i = \frac{V_i}{V} .
\end{equation*}
Similarly, constituents in vascular and cellular compartments must satisfy
\begin{equation*}
\phi_{PV} + \phi_{LV} + \phi_{\sim PLV} = \sigma_v = \frac{V_v}{V}, \qquad \phi_{PC} + \phi_{LC} + \phi_{\sim PLC} = \sigma_c = \frac{V_c}{V} .
\end{equation*}

Constituent volume fractions defined above present a difficulty in the sense that, for example, $\phi_P$, at a point $\bx \in \Omega$ depend on $\sigma_i$ due to saturation condition, and therefore spatial variation of $\sigma_i$ (and similarly $\sigma_v, \sigma_c$) is needed. 
To avoid this difficulty, $\sigma_i, \sigma_v, \sigma_c$ are assumed to be constant (spatially and temporally), 
and 
$\phi_A$ for constituent $A$ in interstitial compartment is defined as $\phi_A = V_A / V_i$ for $A\in \{P, L, \sim PL\}$. Constituents in other two compartments are defined similarly. 
With this, saturation conditions become
\begin{equation*}
\sum_{A \in \{P, L, \sim PL\}} \phi_A = 1, \qquad \sum_{A \in \{PV, LV, \sim PLV\}} \phi_A = 1, \qquad \sum_{A \in \{PC, LC, \sim PLC\}} \phi_A = 1 .
\end{equation*}

\paragraph{Key assumptions}
To derive PDE-based model of HP agents, some simplifying assumptions will be made. The key assumptions are:
\begin{itemize}
\item[(A1)] Cellular compartment and constituents within it are ignored; however, the extension to include this compartment is straight-forward;
\item[(A2)] Tissue domain is treated as homogeneous media and properties such as diffusivities are assumed to be uniform;
\item[(A3)] HP agents in vascular compartment are fixed and given. HP pyruvate in vascular compartment, $\phi_{PV}$, is spatially uniform and varies in time following the gamma probability density function on $\Omega_V$ subdomain of $\Omega$. 
HP lactate in vascular compartment, $\phi_{LV}$, is fixed to zero;
\item[(A4)] Perfusion of HP agents is simulated using volume source terms in the balance of mass of HP agents in interstitium. Perfusion is restricted to subdomain $\Omega_V$. 
Further, exchange of mass is considered one-way, i.e., HP agents perfuse from vascular to interstitial compartment and not in the other direction;
\item[(A5)] The mechanical and thermal effects are ignored. This is a realistic assumption as typically HP-MRI measurements take place over a short time interval ($\approx$ 100 seconds) and mechanical effects over this short interval can be safely ignored; 
\item[(A6)] In the balance of mass of constituents in interstitium, besides the reaction between HP agents and natural loss of signal, agents are assumed to diffuse. Convection effects are ignored. This may not be a realistic assumption as the time scale of measurements suggest that perfusion from capillaries and convection are dominant mechanism for mobility of agents as compared to the diffusion. Future work may address this limitation in the model; and
\item[(A7)] The signal loss due to excitation is assumed to be instantaneous (reasonable assumption considering the time interval of one scan and overall time interval of simulation). 
\end{itemize}

\section{Numerical methods for forward simulation and MI optimization}\label{s:discretization}
In this section, the numerical discretization of the HP-MRI model (mainly HF model) and numerical implementation of the optimization problem to determine optimal design parameters are discussed. 
To solve the LF model presented in \cref{ss:model}, a MATLAB function \texttt{ode45} is employed.  
Numerical discretization of the HF model is discussed next.

\subsection{Discretization of the HF model}
\begin{figure}
\centering
\begin{tabular}{cc}
\includegraphics[clip, trim=35 5 0 45,width=.3\textwidth]{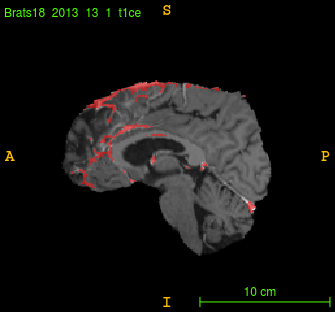} &
\includegraphics[clip, trim=0 0 0 20,width=.3\textwidth]{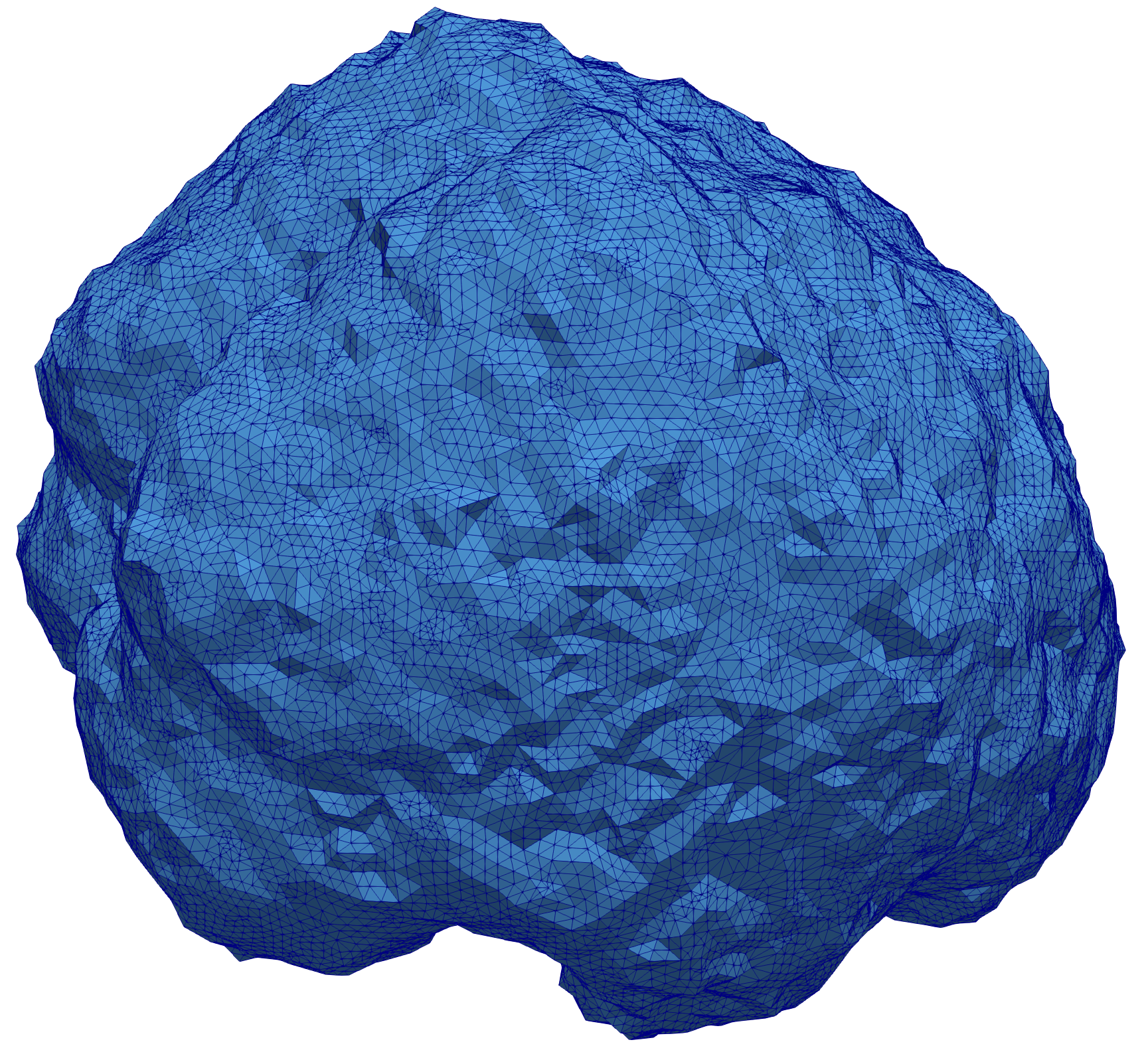}   \\
(a) & (b) \\
\end{tabular}
\caption{ (a) Brain T1 contrast enhanced MRI with segmented vascular (red regions).
The vascular regions represent $\Omega_V$.
 The MRI data consists of 240$\times$240$\times$155 voxels. In (b), a finite
element mesh of the tissue is depicted. }\label{fig:tissue} 
\end{figure}

The HF model is applied to real brain tissue; the sample data is taken from the publicly available BRATS brain MRI datasets~\cite{menze2014multimodal}. 
The vascular domain $\Omega_V$ was extracted from the MRI data in several steps:
First, a four component Gaussian mixture model~\cite{avants2011open} was applied to the 
subtraction image of the difference between the pre- and post-contrast
T1-weighted imaging.
Hessian-based~\cite{frangi1998multiscale} filters were applied to the posterior image class with the
brighted intensity. Otsu thresholding~\cite{otsu1979threshold} is applied to extract the foreground
vessel mask. Finally, a 4$\times$4 axial plane dilation is applied as a post processing step to
oversegment the vessels and ensure smaller vessels were simply connected. 
In \cref{fig:tissue}, the vascular segmentation (a) and the
finite element mesh of the brain tissue (b) are shown.

The coupled PDEs in \eqref{eq:modelHFRep} are discretized using finite element method and a implicit first order time discretization. 
Let $V_h\subset H^1(\Omega)$ denote the finite element space spanned by a second order shape functions over conforming tetrahedral triangulation of $\Omega$. 
Let ${\pyrh^k}_{n}, {\lach^k}_{n}, {\pyrvh^k}_{n}, {\lacvh^k}_{n} \in V_h$ denote the finite element solution at the \kth{n} time step at time $\tau_n = t_k + n\Delta t_k$, where $\Delta t_k$ is a time step size in interval $[t_k, t_{k+1})$ ($\Delta t_k$ can vary with $k$).  
Suppose $\pyrh^{k-1}(t_{k-1}), \lach^{k-1}(t_{k-1}) \in V_h$ are already computed and known. 
Then the initial condition ${\pyrh^k}_{0}, {\lach^k}_{0}$ are applied using \eqref{eq:icHF}:
\begin{equation}
\weakDot{{\pyrh^k}_0}{\testfn}  = \weakDot{\cos(\theta^{k-1}_P) \pyrh^{k-1}(t_{k-1})}{\testfn}, \qquad \weakDot{{\lach^k}_0}{\testfn}  = \weakDot{\cos(\theta^{k-1}_L) \lach^{k-1}(t_{k-1})}{\testfn}, \qquad \forall \testfn \in V_h.
\end{equation}
As discussed in \cref{ss:modelhf}, HP agents in the vascular domain, ${\pyrvh^k}_n, {\lacvh^k}_n$, for all $n=0, 1, ..,N_k$, are fixed and given via relation \eqref{eq:vifvasc}. Suppose solution ${\pyrh^k}_{n-1}, {\lach^k}_{n-1}$ at the \kth{(n-1)} step are known. To determine ${\pyrh^k}_n, {\lach^k}_n$, we consider a implicit discretization as follows (noting that $k_{LP} = 0$ and ${\lacvh^k}_n = 0$ for all $k$ and $n$), for all $\testfn\in V_h$,
\begin{equation}\label{eq:pyrHFdisc}
\weakDot{ \frac{{\pyrh^k}_n - {\pyrh^k}_{n-1}}{\Delta t_k} }{\testfn} + \weakDot{D_P \nabla {\pyrh^k}_n}{\nabla \testfn} - \weakDot{ \left[-\frac{1}{T_{1,P}} - k_{PL} \right]{\pyrh^k}_n}{\pyrh^k} = \weakDot{ L_P {\pyrvh^k}_{n} \chi_{\Omega_{V}} }{\testfn}
\end{equation}
and
\begin{equation}\label{eq:lacHFdisc}
\weakDot{ \frac{{\lach^k}_n - {\lach^k}_{n-1}}{\Delta t_k} }{\testfn} + \weakDot{D_L \nabla {\lach^k}_n}{\nabla \testfn} - \weakDot{-\frac{1}{T_{1,L}} {\lach^k}_{n} - k_{PL}{\pyrh^k}_{n}}{\testfn} = 0.
\end{equation}
Observe that the above two equations can be decoupled by solving \eqref{eq:pyrHFdisc} first and then solving \eqref{eq:lacHFdisc}. 
 
Next, the convergence of numerical discretization is established.

\subsubsection{Time and mesh convergence study}
To compare the solution from different mesh and time steps, finite element solutions of problem over a brain tissue is projected onto a uniform mesh with $N_{h} = 16^3$ cubic elements, denoted by $B_h$, of cubic domain $B$, where $B$ is the smallest cube containing $\Omega$. 
Next, at any simulation time $t$, for each cell $K_i$ in $B_h$, the total volume of HP agents are computed; suppose $Q^i_P(t), Q^i_L(t)$ denotes the total volume of pyruvate and lactate at time $t$ for the \text{i} cell in $B_h$. 
Suppose
$Q^i_{P, 1}, Q^i_{P, 2}$ and $Q^i_{L, 1}, Q^i_{L, 2}$, for all $i$, are results from
two simulations, then, the total error in HP pyruvate and lactate at time $t$ is defined as
\begin{equation}\label{eq:err}
e^{1,2}_P(t) = \frac{1}{\sqrt{|K|}} \sqrt{ \sum_{1 \leq i \leq N_h} |  Q^i_{P,1}(t) - Q^i_{P,2}(t) |^2}, \qquad
e^{1,2}_L(t) = \frac{1}{\sqrt{|K|}} \sqrt{ \sum_{1 \leq i \leq N_h} |  Q^i_{L,1}(t) - Q^i_{L,2}(t) |^2} \,,
\end{equation}
where $|K| = 1042.672$ is the volume of one element in $B_h$. 

The high-fidelity model was solved approximately using FEniCS library \cite{AlnaesBlechta2015a, LoggMardalEtAl2012a}. 
In the convergence results presented below, model parameters are taken from \cref{tab:modelParam} 
while the design parameters are taken as, for $1\leq k \leq N = 30$, 
$\theta^k_P = 35$ and $\theta^k_L = 28$ degrees, respectively 
(these parameters are the OED parameters for SNR = 2, see \cref{fig:snr2}). 
Repetition times are $TR_k = 3$ seconds, for $2\leq k \leq N$. 

\begin{figure}[h]
\centering
\includegraphics[width=0.6\textwidth]{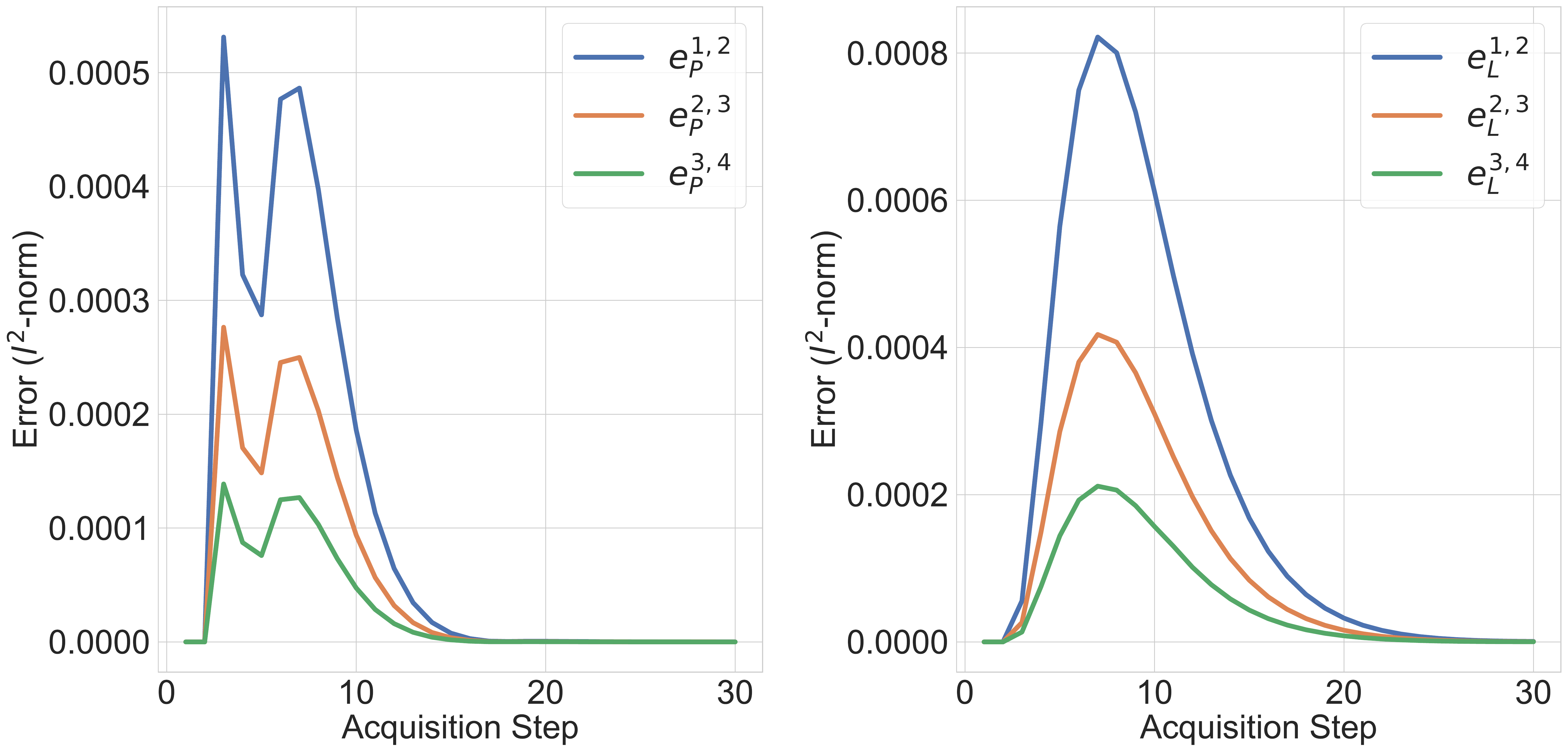}
\caption{Plot of errors for pair of simulations at scan times $t_k = (k-1) * 3$ seconds, $1\leq k \leq 30$. 
Simulation with $\Delta t_k = 0.6$ seconds is labeled as 1 and the one with $\Delta t_k = 0.075$ seconds as 4, 
and the errors for pair of simulations are computed using \eqref{eq:err}. 
The relative error is seen to decrease as time step is decreased. }\label{fig:dtconvg}
\end{figure}

To verify the convergence of solver with decreasing time-step sizes, model was solved with a fixed mesh consisting of $333049$ elements and $68991$ vertices and different time steps $\Delta t_k$. 
Time steps employed in this study are $\Delta t_k = \{0.6, 0.3, 0.15, 0.075\}$ seconds for all $k$. 
Simulation with largest $\Delta t_k = 0.4$ is labeled as 1 and the one with smallest $\Delta t_k = 0.05$ is labeled 4. 
The errors as defined in \eqref{eq:err} for pair of simulations are shown in \cref{fig:dtconvg}. 
The error is seen to decrease consistently with the decreasing time step. 

\begin{figure}[h]
\centering
\includegraphics[width=0.6\textwidth]{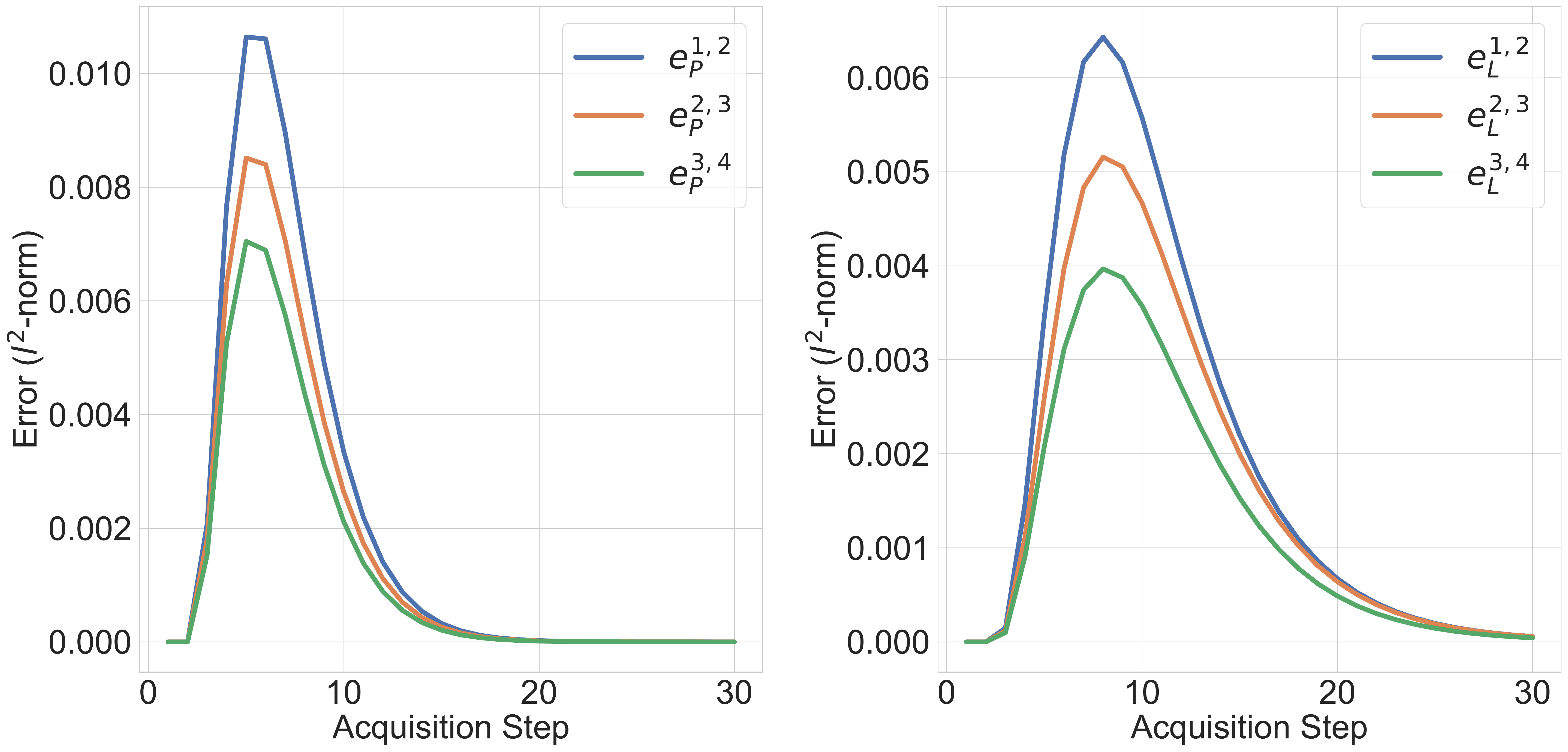}
\caption{Plot of errors for pair of simulations at scan times $t_k = (k-1) * 3$ seconds, $1\leq k \leq 30$. 
Four simulations are carried out each with different mesh keeping the time step fixed to $\Delta t_k = 0.075$ seconds; simulation 1 corresponds to the coarsest mesh and simulation 4 corresponds to the finest mesh. 
The relative errors are seen to decrease as the mesh is refined. }\label{fig:hconvg}
\end{figure}

To ascertain mesh convergence, model was solved with a fixed time step $\Delta t_k = 0.075$ seconds 
and four difference meshes. 
The number of tetrahedral elements and vertices in four meshes employed in this study are: $(239697, 48695)$, $(333049, 68991)$, $(404492, 84991)$, $(523746, 108276)$. 
Results with coarsest mesh (smaller number of elements) is labeled as 1 while the one with finest mesh is labeled as 4. 
The errors for different pair of simulations are shown in \cref{fig:hconvg}. 
The error is seen to decrease as the mesh is refined. 

Based on the time and mesh convergence results, 
to compute ``ground truth", the mesh with $239697$ elements and $48695$ vertices is employed 
and the time step is fixed to $\Delta t_k = 0.15$ seconds.
 
\subsection{Automatic differentiation accelerated optimization for OED calculations}
The auto-differentiation functions of MATLAB were used to calculate gradients
of \eqn{mi3}.
In particular, design parameters $\mathcal{K}$ and state variables $\pyrlac$
were considered as optimization variables
to minimize the objective function \eqn{mi3} with respect to the model
constraints \eqn{eq:model}.
Auto-differentiation provides the derivatives of the objective function and
constraints with respect to this full space formulation. Given the derivatives in
the full space formulation, the reduced space gradient of the 
objective function \eqn{mi3} with respect to the design parameters $\mathcal{K}$
may be calculated using an adjoint solve.

\subsection{Inverse problem to recover rate parameter}
A MATLAB routine \texttt{fmincon} is used to solve the inverse problem of recovering model parameters $\p$ in the LF model \eqref{eq:model} from the data. Signals of pyruvate and lactate at different scans is taken as data.  
As an objective function for the inverse problem, square of the difference between data and model prediction of signals is used. 
Similarly, derivatives from the automatic differentiation feature in the MATLAB are
utilized for numerical optimization.

\end{document}